%% file: setV2017.tex
\documentclass[preprint,prd,showpacs,superscriptaddress,amssymb,amsmath,preprintnumbers]{revtex4}
\def\DLB{\Delta\!L\!B}
\def\LB{L\!B}

\usepackage{amsmath}
\usepackage{amssymb}
\usepackage{bm}
\usepackage{dcolumn}
\usepackage{graphicx}
\usepackage{longtable}
\usepackage{afterpage}

\begin{document}

\date{\today}


\title{%
Revised and Improved Value of the QED Tenth-Order Electron Anomalous Magnetic Moment 
}


\author{Tatsumi Aoyama}
\affiliation{Yukawa Institute for Theoretical Physics, Kyoto University, Kyoto, Japan 606-8502}
\affiliation{Nishina Center, RIKEN, Wako, Japan 351-0198 }


\author{Toichiro Kinoshita}

\affiliation{Laboratory for Elementary Particle Physics, Cornell University, Ithaca, New York, 14853, U.S.A. }
\affiliation{Amherst Center for Fundamental Interactions, Department of Physics, University of Massachusetts, Amherst, MA, 01003, U.S.A.}

\author{Makiko Nio}
\affiliation{Nishina Center, RIKEN, Wako, Japan 351-0198 }

\begin{abstract}
In order to improve the theoretical prediction of the electron anomalous magnetic moment $a_e$ 
we have carried out a new numerical evaluation of the 389 integrals of Set~V, which represent 6,354 Feynman vertex diagrams without lepton loops.
During this work, we found that
one of the integrals, called $X024$, was given a wrong value
in the previous calculation
due to an incorrect assignment of integration variables.
The correction of this error causes a shift of $-1.25$ to the Set~V contribution, and hence to
the tenth-order universal (i.e., mass-independent) term $ A_1^{(10)}$.  The previous evaluation of 
all other 388 integrals is free from errors
and consistent with the new evaluation.
Combining the new and the old (excluding $X024$) calculations
statistically, we obtain 
$7.606~(192) (\alpha/\pi)^5$
as the best estimate
of the Set~V contribution.  
 Including the contribution
of the diagrams with fermion loops, the improved tenth-order universal term  becomes
$A_1^{(10)}=6.678~(192)$. 
Adding hadronic and electroweak contributions leads to the theoretical prediction 
$a_e (\text{theory}) =1~159~652~182.032~(720)\times 10^{-12}$. 
From this and the best measurement of $a_e$, we obtain the
inverse fine-structure constant $\alpha^{-1}(a_e) = 137.035~999~1491~(331)$.
The theoretical prediction of the muon anomalous magnetic moment 
is also affected by
the update of QED contribution and the new value of $\alpha$, 
but  the shift is much
smaller than the theoretical uncertainty. 
\end{abstract}



\maketitle

\section{Introduction and Summary}
\label{sec:intro}

In 1947 the electron magnetic moment anomaly $a_e = (g - 2)/2$ was discovered in an atomic physics experiment \cite{Kusch:1947},
which was soon understood as the effect of radiative correction by the
newly formulated quantum electrodynamics (QED) \cite{Schwinger:1948iu}.
Since then comparison of measurement and
theory of $a_e$ has provided more and more stringent test of 
QED and the standard model (SM) of elementary 
particles. 

The most accurate measurement of $a_e$ thus far has been carried out by the Harvard group  
using a cylindrical Penning trap
\cite{Hanneke:2008tm,Hanneke:2010au}:  
\begin{eqnarray}
a_e(\text{HV08})= 1~159~652~180.73~ (28) \times 10^{-12} ~~~[0.24 \text{ppb}].
~
\label{eq:aeHV08}
\end{eqnarray}
The precision of this value is fifteen times higher than that of the pioneering work by the group at the University of Washington \cite{VanDyck:1987ay}.
Further improvements for the electron and positron measurements are currently being prepared 
by the Harvard group \cite{Hoogerheide:2014mna}.

To test the validity of the theory of $a_e$, it must be evaluated to match
the precision of the measurement \eqref{eq:aeHV08}.
The dominant contribution comes from QED, while 
at such a precision,  the SM contribution can no longer be ignored. Thus we can write 
\begin{equation}
a_e = a_e({\rm QED}) + a_e({\rm Hadron}) + a_e({\rm Weak}).
\label{eq:aeformula}
\end{equation} 
The QED contribution can be expressed further, by taking the heavier leptons ($\mu$ and $\tau$) into account, as
\begin{equation}
a_e({\rm QED}) = A_1 + A_2(m_e/m_\mu) + A_2(m_e/m_\tau) + A_3(m_e/m_\mu, m_e/m_\tau).
\end{equation} 
Note that the mass-dependence appears in the form of mass ratio because $a_e$ is dimensionless.
All four terms are expressed in the perturbation series of the fine-structure constant $\alpha$
\begin{equation}
A_n = \sum_{i=1,2, \cdots} \left( \frac{\alpha}{\pi} \right )^{i}  A_n^{(2i )} .
\end{equation}  
Since the electron is the lightest lepton, contributions from heavier particles are
suppressed and tiny, although not negligible.

The QED contribution involving heavy leptons are known with sufficient precision.
The muon and tau-lepton contributions $A_2$ and $A_3$ of $a_e$ up to eighth order 
have been calculated both numerically and analytically,
with a good agreement  with each other
\cite{Elend:1966a,Samuel:1990qf,Li:1992xf,Laporta:1992pa,Laporta:1993ju,Passera:2006gc,
Kinoshita:2005sm,Kataev:2012kn,Kurz:2013exa}.
The tenth-order mass-dependent contribution $A_2^{(10)}(m_e/m_\mu)$ 
has been evaluated numerically \cite{Aoyama:2008gy,Aoyama:2008hz,Aoyama:2010yt,Aoyama:2010pk,
Aoyama:2010zp,Aoyama:2011rm,Aoyama:2011zy,Aoyama:2011dy,Aoyama:2012fc, ae10:PRL}. 
Some of the tenth-order diagrams 
have been independently calculated and checked \cite{Baikov:2013ula}. 
The tau-lepton contribution to the tenth-order term is 
currently negligible compared to the $A_1$ term
since it is suppressed by the factor $(m_e/m_\tau)^2$ 
and contributes to $a_e$ no more than $\mathcal{O}(10^{-18}) $.
Summing all mass-dependent terms, we obtain 
\begin{equation}
a_e(\text{QED:mass-dependent}) = 2.747~5719~ (13) \times 10^{-12},
\label{eq:mass-depend}
\end{equation}
where the uncertainty comes from the tau-electron mass ratio. The uncertainty
due to the muon-electron mass ratio is one order of magnitude smaller and is about $0.13 \times 10^{-18}$.

To compare the theory with the experiment, it is of course necessary to evaluate the mass-independent term $A_1$ 
up to tenth order of perturbation theory,  since  $(\alpha/\pi)^5 \sim 0.07 \times 10^{-12}$.
The second-, fourth-, and sixth-order terms were calculated analytically 
 \cite{Schwinger:1948iu,Petermann:1957,Sommerfield:1958,Laporta:1996mq}
or by numerical or semi-analytical means \cite{Kinoshita:1995ym, Melnikov:1999xp}:
\begin{eqnarray}
A_1^{(2)} &=& 0.5, \\
A_1^{(4)} &=&  -0.328~478~965~579~193\cdots ,     \\
A_1^{(6)} &=&   1.181~241~456~587~\cdots .   
\end{eqnarray} 

Recently, Laporta has reported a highly precise value 
of the eighth-order term $A_1^{(8)}$, 
with an accuracy of 1100 digits, 
after twenty years of persistent research \cite{Laporta:2017okg}.  
Thus the uncertainty due to the eighth-order term has been completely eliminated
by his outstanding work. It took 36 years since the preliminary value 
$A_1^{(8)}=-0.8~(2.5)$ was reported \cite{Kinoshita:1981vs}.
For the purpose of this article it is sufficient to list the first ten digits of Laporta's result:
\begin{equation}
A_1^{(8)}[\mbox{semi-analytic}] =  -1.912~245~764 \cdots  ,
\label{eq:Laporta8th}
\end{equation}
which confirms the validity of the earlier numerical evaluation \cite{Aoyama:2014sxa} 
\begin{equation}
A_1^{(8)}[\mbox{numerical}] = -1.912~98~ (84) .
\label{eq:our8th}
\end{equation}
Although it is less accurate than \eqref{eq:our8th},
another semi-analytic result $A_1^{(8)}=-1.87~(12)$ 
given in Ref.~\cite{Marquard:2017iib} is consistent with
both of preceding results \eqref{eq:Laporta8th} and \eqref{eq:our8th}.
The contribution to $A_1^{(8)}$ from 
the 518 vertex diagrams without a fermion loop has also been independently 
cross-checked by using numerical means\cite{Volkov:2017xaq}.

The tenth-order mass-independent term $A_1^{(10)}$ is thus
 the only significant QED contribution
which has not been verified by independent calculations.
It has a contribution from 12,672
vertex-type Feynman diagrams.
Of these diagrams those that are dominant and the hardest to evaluate belong to  Set~V,
which consists of 6,354 vertex diagrams without a fermion loop.
Two of these vertex diagrams 
have been evaluated by other means thus far.  Their values are given in Refs.~\cite{Caffo:1978mg, Volkov:2017xaq}.
We have compressed all 6,354 vertex diagrams to 389 integrals 
by certain algebraic manipulation. 
The preliminary value of the contribution of Set~V obtained by numerical integration by VEGAS
was \cite{Lepage:1977sw, Aoyama:2014sxa} 
%
\begin{equation}
A_1^{(10)}[\mbox{Set V}: 2015] = 8.723 ~(336).
\label{eq:SetV2015}
\end{equation}
In order to improve this further,
we have re-evaluated these 389 integrals using independent sets
of integration variables.  The result is 
\begin{equation}
A_1^{(10)}[\mbox{Set V}: 2017]  = 7.791~(264),
\label{eq:SetV2017}
\end{equation}
which disagrees with 
\eqref{eq:SetV2015} by $-0.93$.

\begin{figure}[t]
\vspace{-3cm}
\resizebox{12cm}{!}{\includegraphics{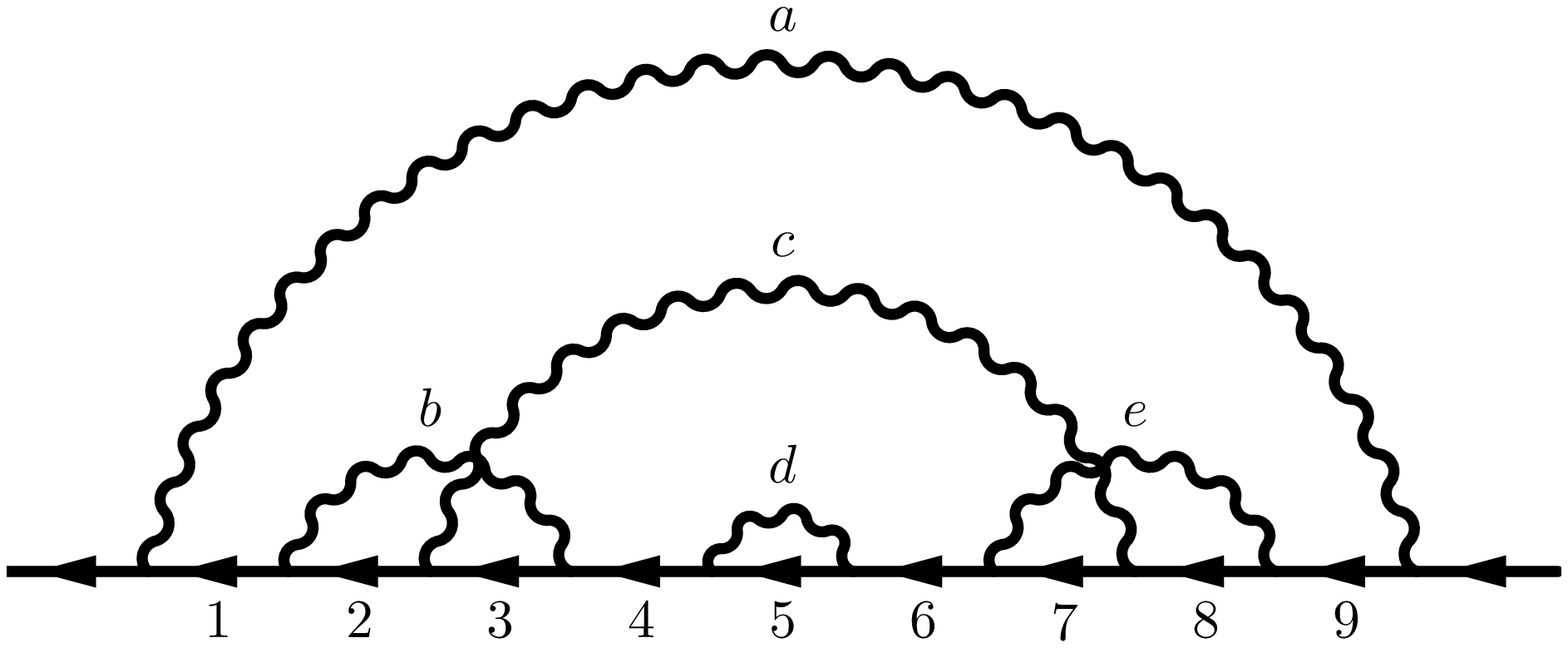}}
\vspace{-6cm}
\caption{
\label{fig:X024}
Self-energy-like diagram $X024$($abcbddecea$).  The straight and wavy lines represent fermion and
photon propagators, respectively.  Indices assigned to the fermion lines
are $1,2,\cdots, 9$ from left to right, and those to the photon lines are $a,b,\cdots,e$.
The nine vertex diagrams related to this self-energy-like diagram are obtained by inserting an
external photon vertex in each of the nine fermion lines. 
}
\end{figure}

This discrepancy arises mainly from the integral $X024$ 
expressing the contribution from  Fig.~\ref{fig:X024}, which
represents the sum of nine vertex diagrams.
During the new evaluation
we found a programming error in the previous evaluation of $X024$.
When the error was corrected, 
its numerical value shifted from $-6.0902~(246)$ to $-7.3480~(139)$. The difference of $-1.26$ 
accounts for almost all the difference between the values \eqref{eq:SetV2015} and \eqref{eq:SetV2017}.
If this correction of $X024$ is added to the old result \eqref{eq:SetV2015}, we obtain
\begin{equation}
A_1^{(10)}[\mbox{Set V}:2015{\rm corrected}] = 7.465~(335),
\label{eq:SetV2015corrected}
\end{equation}
which is consistent with the new value of Set~V given in
Eq.~\eqref{eq:SetV2017}.
The details of the origin of the error in the $X024$ integral are discussed in Sec.~\ref{sec:X024 error corrected}.

Since the new calculation \eqref{eq:SetV2017} is independent of
the old one, two calculations can be statistically combined
for each integral. The best estimate for  Set~V then becomes
\begin{equation}
A_1^{(10)}[\mbox{Set V}] = 7.606~(192)~.
\label{eq:SetVbest}
\end{equation}
Adding the contribution from the 6,318 diagrams with fermion loops \cite{Aoyama:2008gy,Aoyama:2008hz,Aoyama:2010yt,Aoyama:2010pk,
Aoyama:2010zp,Aoyama:2011rm,Aoyama:2011zy,Aoyama:2011dy,Aoyama:2012fc, ae10:PRL}
\begin{equation}
A_1^{(10)}[\mbox{Set I--IV,VI}] =  -0.930~42~(361)
\end{equation} 
to \eqref{eq:SetVbest}, 
we obtain an updated value of the tenth-order mass-independent contribution:
\begin{equation}
A_1^{(10)}= 6.675~(192)~,
\end{equation}
where the uncertainty comes entirely from the numerical integration
of Set~V  and 
is reduced by 43\% compared to that in \eqref{eq:SetV2015}.
This is the main result of our paper.

The contributions of the electroweak interaction and the hadronic interaction have been updated recently 
\cite{Jegerlehner:2017zsb}
including new hadronic measurements \cite{BESIII:2015orh,Anashin:2016hmv}:
\begin{eqnarray}
a_e({\rm Weak}) &=& 0.030~53 ~(23) \times 10^{-12}, \nonumber \\
a_e({\rm Hadron}) &=& \{ 1.8490~(108) -0.2213~(12) +0.0280~(2) + 0.037~(5) \} \times 10^{-12} \nonumber \\
                               &=& 1.6927~(120) \times 10^{-12},
\label{eq:hadweakJegerlehner}
\end{eqnarray} 
respectively, where the hadronic contribution consists of  the leading-order (LO), next-to-leading-order (NLO), and next-to-next-to-leading-order (NNLO) vacuum-polarization (VP)  contributions 
and the hadronic light-by-light scattering contribution from left to right.
The combined uncertainty  of $a_e({\rm Hadron})$ is the one given in Eq.~(5) of Ref.~\cite{Jegerlehner:2017zsb}.



It is noted that the same spectral function is used 
to obtain the LO-, NLO-, and NNLO-VP contributions, and their 
systematic uncertainties are correlated, as pointed out in Refs.~\cite{Nomura:2012sb, BarryTaylor2017}. 
The correlation should be taken into account to derive the combined 
uncertainty \eqref{eq:hadweakJegerlehner}.

Summing up all the contributions of SM,  we obtain the theoretical prediction for  $a_e$ as
\begin{equation}
a_e(\text{theory}) = 1~159~652~182.032~(13)(12)(720) \times 10^{-12},
\label{eq:aetheory}
\end{equation}
where the first and second uncertainties are due to the tenth-order QED and the hadronic
corrections, respectively. 
The uncertainty due to the mass ratios of tau or muon to electron is currently negligible. 
The third and largest uncertainty comes from the fine-structure constant $\alpha$.
Here, we used the latest value of $\alpha$ \cite{Mohr:2015ccw,Mohr:2017}, determined
from the recoil measurement of the Rb atom $h/M_{\text{Rb}}$ \cite{Bouchendira:2010es} combined with
the relative atomic mass of the electron $A(e)$ from the $g$-factor of the bound electron \cite{Mohr:2015ccw},
the relative atomic mass of the Rb atom $A(\text{Rb})$ \cite{AtomicMassRb2012a,AtomicMassRb2012b}, and 
the Rydberg constant $R_\infty$ \cite{Mohr:2015ccw}:
\begin{equation}
\alpha^{-1}(\text{Rb:2016}) =137.035~998~995~(85)~,
\label{eq:alinvRb2016}
\end{equation}
which replaces  the value given in Ref.~\cite{Mohr:2012tt}, $\alpha^{-1}(\text{Rb:2010}) = 137.035~999~049~(90)$.
%
%

The new theoretical value of $a_e$ is greater by $0.38 \times 10^{-12} $
than that of Eq.~(16) of Ref.~\cite{Aoyama:2014sxa}. 
The corrected and updated  $A_1^{(10)}$ adds  $-0.08 \times 10^{-12} $ to $a_e$,
the near-exact value of $A_1^{(8)}$  increases $a_e$ by $ 0.02 \times 10^{-12}$,
and the new value $\alpha(\text{Rb})$ increases $a_e$ by $ 0.45 \times 10^{-12}$.
The shift due to the new values of the electroweak and hadronic contributions is $-0.01 \times 10^{-12}$. 
The difference between experiment \eqref{eq:aeHV08} and theory \eqref{eq:aetheory} is thus
\begin{equation}
a_e(\text{HV08}) - a_e(\text{theory}) = (-1.30 \pm 0.77 )\times 10^{-12} .
\label{eq:ae_exp-theory}
\end{equation}
 
 If we assume that the theory of $a_e$ is correct,  by equating the formula Eq.~\eqref{eq:aetheory}
 to the measured value Eq.~\eqref{eq:aeHV08}, we obtain an  $\alpha$ 
which is more precise than that of \eqref{eq:alinvRb2016}:
 \begin{equation}
 \alpha^{-1}(a_e:2017) = 137.035~999~1491~(15)(14)(330),
 \label{eq:alinv_ae}
 \end{equation}
where the uncertainties come from the tenth-order QED, the hadronic correction,
and the experiment. The shift from the previous value given in  Eq.~(18) of  Ref.~\cite{Aoyama:2014sxa}
amounts to $ -0.87 \times 10^{-8}$ and  is due to the new values of $A_1^{(8)}$, $A_1^{(10)}$, and the electroweak
and hadronic corrections.
The difference between two determinations 
\eqref{eq:alinv_ae}  and \eqref{eq:alinvRb2016}
of $\alpha$ is
\begin{equation}
\alpha^{-1}(a_e:2017) - \alpha^{-1}(\text{Rb:2016}) = ( 0.155 \pm  0.091 ) \times 10^{-6}.
\end{equation}

Since the updated QED contributions of $A_1^{(8)}$ and $A_1^{(10)}$ are universal for
any lepton species, the theoretical prediction of the muon anomalous magnetic moment ($a_\mu$) 
should also be changed. The new value of the fine-structure constant $\alpha$ derived from $a_e$  also
causes a small shift in  $a_\mu$. The total shift caused by them is, however, far smaller than  the
current theoretical uncertainty and has no significant influence  on comparison of theory
and experiment of $a_\mu$.
The details of the updates of $a_\mu$ is given in Appendix \ref{sec:appendix_amu}. 

\section{Preparation of Set~V  for Numerical Evaluation}
\label{sec:SetV}

QED is renormalizable.  But individual Feynman diagrams are divergent.
For numerical integration, it is absolutely necessary that integrals do not contain any divergence.
This means that all divergences must be removed before integration is carried out.

Another problem with Set~V, which
consists of 6,354 tenth-order vertex diagrams without a fermion loop,
is its huge size. 
It should be difficult to achieve high precision in the numerical 
evaluation 
because of accumulation of uncertainties of individual 
diagrams, even though each diagram were evaluated precisely enough. 
Thus it is highly desirable to reduce the number of independent integrals as much as possible.
One way to achieve such reduction is to sum up 9 vertex diagrams obtained by inserting an external photon vertex 
in each of 9 fermion lines in a tenth-order electron self-energy diagram which does not contain a closed fermion loop.
Let $\Sigma (p)$ be such a self-energy diagram
and $\Lambda^{\nu}(p,q)$ the sum of 9 vertex diagrams related to 
$\Sigma (p)$.
Then, with the help of an equation derived from the Ward-Takahashi identity, one finds 
%
%
\begin{equation}
\Lambda^{\nu}(p,q) \approx q_{\mu} \left [  \frac{\partial \Lambda^\mu (p,q)}{\partial q_\nu} \right ]_{q=0}
- \frac{\partial \Sigma (p) }{\partial p_\nu}~
\label{eq:WTsum}
\end{equation}
in the small $q$ limit.
The sum of all vertex diagrams of Set~V can then be represented by 706 quantities of the form given on the right hand side of Eq.~\eqref{eq:WTsum}, 
which we shall call ``self-energy-like'' diagram.
Taking the time-reversal symmetry into account, we can reduce it further to 389 self-energy-like diagrams which represent all vertex
diagrams of Set~V.  


We assign Feynman parameters 
$z_1,\cdots, z_9$ to nine fermion lines 
and $z_a, \cdots, z_e$ to five photon lines from the left to the right.
They are subject to the  constraint 
$z_1 + \cdots + z_9 +z_a + \cdots +z_e=1$.   

Each self-energy-like  diagram is then represented by a sequence of ten vertices on the fermion line
labeled by the photon Feynman parameters. For instance, $X024$ in Fig.~\ref{fig:X024} is represented by the sequence
``$abcbddecea$''.

The integral 
for a self-energy-like diagram $\mathcal{G}$ 
is defined in the Feynman parameter space and has a form
\begin{align}
M_\mathcal{G}= \left ( \frac{-1}{4} \right )^5 4 ! & \int(dz)_\mathcal{G}
                             \left [   \frac{1}{4}   \left ( \frac{E_0+C_0}{U^2 V^4}  \right .
                                                               + \frac{E_1+C_1}{U^3 V^3} 
                                                               +   \cdots
                                                                +   \frac{E_3+C_3}{U^5 V}
                                                                                                                         \right ) 
            \nonumber \\              
                                & +    \left .           \left ( \frac{N_0+Z_0}{U^2 V^5}
                                                               + \frac{N_1+Z_1}{U^3 V^4} 
                                                               + \cdots
                                                                      + \frac{N_4+Z_4}{U^6 V} 
                                                      \right )
                               \right ]~,
\end{align}
where $(dz)_\mathcal{G} = dz_1 ...dz_9 dz_a ...dz_e \delta (1 - z_1 \cdots - z_9 - z_a \cdots - z_e)$, all $z_i$ being non-negative.
$E_n$ and $C_n$ terms  and $N_n$ and $Z_n$ terms are projected out
from the first and second terms of the right-hand-side of Eq.~\eqref{eq:WTsum}, respectively.
All functions $E_n, C_n, N_n, Z_n, V, U$ are expressed in terms of the Feynman parameters 
and the \textit{building blocks} $A_i, B_{ij}, C_{ij}$ for $i,j =1,2,\cdots,9$, which are polynomials of 
Feynman parameters. 
The detailed definitions of  these functions are given in Refs.~\cite{Cvitanovic:1974uf, Kinoshita:1990, Aoyama:2005kf}.

The bare amplitude $M_\mathcal{G}$  is inherently ultraviolet (UV) divergent. If  the diagram $\mathcal{G}$ 
possesses a self-energy subdiagram, it 
suffers from infrared (IR) divergence as well.
Though we do not explicitly write them, the amplitude $M_\mathcal{G}$ is regularized by
the Feynman cut-off for UV divergence and the small photon mass for IR divergence. Both regulators
are safely and harmlessly removed after  the finite integral is constructed with the bare amplitude
$M_\mathcal{G}$ and the corresponding UV and IR counterterms.

The UV divergence emerges with a self-energy or vertex subdiagram, and the entire divergence structure is obtained in the form of Zimmermann's forests \cite{Zimmermann:1969}. 
The $\bm{K}$-operation, which extracts the most UV-divergent part of a subdiagram,  is applied  to $M_\mathcal{G}$
according to Zimmermann's forest formula, and the UV subtraction terms are generated \cite{Cvitanovic:1974sv,Aoyama:2005kf}.
For IR divergences, the divergent structure is 
given by the \textit{annotated} forests of self-energy subdiagrams 
to which are assigned either magnetic moment or
self-mass properties. 
Either the $\bm{I}$- or $\bm{R}$-subtraction operation is applied to $M_\mathcal{G}$ according to the annotation to generate 
the IR subtraction terms \cite{Aoyama:2007bs, Aoyama:2014sxa}. UV divergence arising in the IR subtraction terms are
further subtracted  by applying the $\bm{K}$-operation to them.

The finite magnetic moment amplitude $\Delta M_\mathcal{G}$ 
as the output of these operations on the diagram $M_\mathcal{G}$,
made UV-finite by $\bm{K}$-operation
and IR-finite by $\bm{R}$- and/or $\bm{I}$-operations,
is thus symbolically written as 
\begin{eqnarray}
  \Delta M_\mathcal{G} &= &M_\mathcal{G}
  \nonumber \\ 
  &+ &\sum_\text{forests} (- \bm{K}_{S_i})   \cdots (-\bm{K}_{S_j}) M_\mathcal{G}
  \nonumber \\                     
  & +& \sum_\text{\textit{annotated} forests} (-\bm{I}_{S_i} ) \cdots  (-\bm{I}_{S_j} ) 
  (-\bm{R}_{S_k})  \cdots   (-\bm{R}_{S_l} ) M_\mathcal{G}.
\end{eqnarray}
The sum  over all diagrams 
is denoted as $\Delta M_{10} $
\begin{equation}
\Delta M_{10}= \sum_{\mathcal{G}= X001}^{ X389} \Delta M_\mathcal{G},
\end{equation}
where the time-reversal symmetric factor 2 is included in the definition of $\Delta M_\mathcal{G}$.

The procedure to generate the finite amplitude can be handled in an automated 
way by the code-generating program {\sc{gencode}\it{N}}. 
It takes the one-line expression of a diagram $\mathcal{G}$ using  a sequence of photon labels 
on the fermion line and
creates the form of the bare amplitude. The generator then 
identifies the divergent subdiagrams and their forests, 
and constructs the corresponding UV and IR subtraction terms. 
Finally, the finite amplitude ready 
to be integrated by the numerical integration routine is 
generated \cite{Aoyama:2007bs, Aoyama:2014sxa}.
Numerical integration of individual $\Delta M_\mathcal{G}$ is described in the next section.

\section{Numerical Integration of $\Delta M_{10}$ by VEGAS}
\label{sec:evaluation by vegas}

Because of the linear constraint imposed on the 14 Feynman parameters, the integration domain $(dz)_\mathcal{G}$ is a
13 dimensional hyperplane.  
For a diagram which has no self-energy subdiagrams, 
the 14 Feynman parameters of the integrand are mapped onto a 13 dimensional unit hypercube 
that is the integration domain of the integration algorithm VEGAS.
%
%
There are many ways to realize the mapping 
from the 14 Feynman parameters to the 13 integration variables.
Analytically any choices of mapping, of course, give the same value.

For a diagram with a self-energy subdiagram, we can reduce the
number of integration variables.  After projecting out the 
magnetic moment contribution from a self-energy-like diagram, 
the integrand depends only on the sum of 
the adjacent Feynman parameters $z_{i-1}+z_{j+1}$, 
if the diagram contains a self-energy subdiagram $S_{ij}$ containing the fermion lines $\{i, i+1, \cdots, j-1,j\}$
from the left to the right.
In the case of $X024$ ($abcbddecea$), it contains two self-energy subdiagrams $S_{28}$ and $S_{55}$ so that the integrand of $X024$
depends  on the sums $z_{19} \equiv z_1 + z_9$ and $z_{46} \equiv z_4 + z_6$. 
The $X024$ can  then be 
numerically evaluated on a 11-dimensional hypercube instead of a 13-dimensional one.

Such reduction of independent variables is related to the loop topology of
the Feynman diagram. 
The adjacent fermion lines attached to a self-energy subdiagram can be regarded topologically as the same line.
Since  they are determined by and only by the loop topology of a diagram,  
the building blocks $B_{i-1, k}$ and $B_{j+1, k}$  for any $k$
have exactly the same expression if the indices $i-1$ and $j+1$ belong to the adjacent fermion lines
of the same self-energy subdiagram $S_{ij}$.
The scalar currents $A_i$ of the adjacent fermion lines are also 
exactly the same because the vanishing limit of the transfer momentum ($q\rightarrow 0$) is 
taken when the magnetic moment contribution is projected out.
For the case of $X024$, the adjacent fermion lines $z_{1}$
and $z_{9}$ lead to $B_{1k}=B_{k9}(=B_{k1}=B_{9k})$ for
any $k$ of the fermion and photon lines. For the scalar currents, $A_1 = A_9$. 
Implementation to these features into {\sc gencode}{\it N} is straightforward.
The resulting integrand becomes much shorter, almost halved in many cases, 
that enables us to reduce the computational cost significantly.

Among the 389 self-energy-like diagrams of Set~V, 
the 254 diagrams have at least one self-energy subdiagram. 
They are grouped as $X\!B$, and have been evaluated with dimensions from 9 to 12, depending on the numbers of self-energy subdiagrams involved.
The remaining 135 diagrams that are grouped as $X\!L$ receive only vertex corrections, and have been evaluated with 13 dimensions. 

An $X\!B$ integrand has at least one IR subtraction term.
The cancellation of the IR divergence within the integrand is in the form of the inverse power law of the vanishing photon mass.
Some of the $X\!B$ integrals suffer from very severe round-off errors 
for finite numerical precision, 
and the VEGAS integration breaks down while the integration is iterated. 
The simplest solution to this \textit{digit-deficiency} problem is
to use an extended precision for real numbers. 
In order to accelerate the convergence of numerical calculation on a computer, 
an algorithm that realizes double-double ({\sc dd}) precision \cite{Hida:2001algorithms} has been adopted.
The arrayed version of the {\sc dd} library was prepared by one of us (T.A.) and used for productive execution. 
The 253 of 254 $X\!B$ integrals have been evaluated with the {\sc dd} precision in the entire integration domain. 

The diagram $X008$ ($abbccddeea$) suffers from the most severe IR cancellation, 
and needs the quadruple-double ({\sc qd}) precision for real numbers \cite{Hida:2001algorithms} in some part of the integration domain. 
We divided the entire integration domain of $X008$ into four, 
and applied the {\sc qd} precision for the most dangerous but narrow region. 
The {\sc dd} precision is used for the regions including the dangerous edges. 
In the remaining wide region, the shape of the integrand $X008$ is relatively 
smooth, and the double precision is sufficient to carry out the integration. 
The same division rule were also adopted for the previous calculation.
We  moved the borders between four regions slightly in the new calculation.

Unlike IR divergences, UV divergence is in powers of logarithms at most. 
The double precision for real numbers implemented in a standard hardware is sufficient to handle the UV cancellation. 
The $X\!L$ integrals involve no IR subtraction terms and have only UV subtraction terms.
They were therefore evaluated in double precision.

It is found that the elapsed wall-clock time needed for VEGAS 
to perform one iteration of an integration 
strongly depends on the array parameter of the arrayed {\sc dd} library. 
After many trial runs, we determined the array parameters for 
computers of three different architectures
to make numerical computation of Set~V as fast as possible. 

We follow the next procedure for a VEGAS integration: 
We start from a flat grid structure. 
The grid is adjusted automatically after each iteration 
with $10^7$ sampling points, that is iterated for 50 times. 
Then the number of sampling points is increased to $10^8$ in several steps, 
and additional 50 iterations are carried out. 
After a good 
grid structure is formed, the productive calculation 
starts with sampling points $2.56 \times 10^9$ or $4 \times 10^9$ 
per iteration, depending on the difficulty of the integration. 
The iteration is repeated at least 2 times, 
until the estimated uncertainty of VEGAS for the integration result 
decreases to less than $0.025$. 

The twelve integrals showing slow convergence and larger uncertainties 
were evaluated once more  with different mappings on a new computer system,
HOKUSAI-BigWaterfall, 
which is almost twice faster than other two systems
and allows us to increase the number of sampling points to 
$1.6 \times 10^{10}$.
So far approximately $3.3 \times 10^7  \text{core}\cdot\text{hours}$ 
of the computer resources 
have been dedicated 
to obtain the 389 new values listed in Table~\ref{table:X001}.
All 389 integrals of new calculation have achieved the required precision and
every uncertainties have been reduced to less than $0.022$.
After consistency between new and old calculations were checked for each integral,
we combined two results except for the $X024$ integral.  
Finally, for every integral, the combined uncertainty is 
reduced to less than $0.018$. 

\section{Mapping of Feynman parameters}
\label{sec:mapping}

The convergence speed of the VEGAS integration of a Set~V integral
does  
depend on a choice of mapping of the Feynman parameters to
the integration variables, 
although analytically any choices of mappings should give the same value.
The shape of an integrand looks very different if a mapping is different.
A mapping should be chosen 
so that the shape of an integrand is 
as flat as possible in the entire integration domain. 
In addition, the results of VEGAS numerical integration with different mappings
can be regarded  as independent of each other. 

In general, an integrand of the higher order perturbation theory of the
anomalous magnetic moment 
has very sharp peaks where UV or IR cancellation occurs. 
Thus,  
a mapping must be selected  in such a way that as small a number of 
sharp peaks as possible are formed as seen by the VEGAS integration variables.
There is no definite rule to select the best mapping, 
but an integral shows better behavior and 
faster convergence against the VEGAS iterations 
if its mapping reflects the structure of divergent subdiagrams.
The mappings of the previous calculation in Ref.~\cite{Aoyama:2014sxa} 
and also in this work were
chosen according to this policy. 
When a mapping is properly chosen, singular regions are concentrated 
at the edges of only a few integration variables. 
This means that the sharp peaks spread out in the surface volumes of 
the whole integration domain.
We further apply the power-law stretches to these edges 
before performing numerical integration in order to accelerate convergence
\cite{Kinoshita:1990}.

For the new calculation we have rewritten the mappings of all 389 integrals 
to different but ``better" ones. Recalculation of the Set~V integrals
with different mappings provides us with a useful check of reliability of 
the VEGAS integration results.
Since the choice of mapping is not deterministic, it was not automated 
in the earlier version of {\sc gencode}{\it N}, and a default thirteen-dimensional 
mapping was commonly used for all integrals of Set~V. 
(The latest version of {\sc gencode}{\it N} writes down a mapping according to 
given instructions.) 
The mapping part occupies 30 lines of a {\sc FORTRAN} code within about 100,000 
lines of the integrand and it is the only part modified by the human hand before
the numerical integration is carried out.

Even though the execution time is limited, the consistent results of
the numerical integration with two different ``better" mappings
indicate that the VEGAS integration result
should be reliable.

\section{Mapping error in $X024$ corrected}
\label{sec:X024 error corrected}

We have integrated all 389 integrals of Set~V according to the scheme outlined above.
A preliminary result was reported in Ref.~\cite{Aoyama:2014sxa}.
In order to improve it further we have carried out a new and independent numerical evaluation
of the integrals of Set~V.
During this work we discovered that one of the integrals, called $X024$,
was given a wrong value in the previous calculation due to an incorrect mapping of Feynman parameters onto the VEGAS integration variables.
The 14 Feynman parameters of $X024$ subject to a linear constraint can  be mapped onto an 11-dimensional unit hypercube because of the 
presence of two self-energy subdiagrams.

The $X024$ has a vertex subdiagram $S_{78}$ consisting
of the fermion lines 7 and 8 and the photon line $e$. 
It causes UV divergence
when the sum $z_{78e}=z_7+z_8+z_e$ tends to zero. Though its divergence
is canceled by a UV counterterm, the integrand shows a very sharp peak near
$z_{78e}=0$.
So, we assigned one of the VEGAS integration variables $q(i), i=1,2,\cdots,$ 
to the sum $z_{78e}$ as $z_{78e}\propto q(9)$ 
that was further mapped as $z_{78}=z_{78e}\times q(10)$ and $z_e=z_{78e}-z_{78}$.
We were supposed to divide $z_{78}$ into $z_7$ and $z_8$ using the eleventh integration variable such that $z_7=z_{78}\times q(11)$ and $z_8=z_{78} - z_7$.
Instead, we treated $z_{78}$ as though the lines 7 and 8 were the adjacent lines of a self-energy subdiagram, 
and assigned the halved value to $z_7=z_{78}/2$. 
Thus the integral was evaluated with 10 dimensions.
This resulted in the wrong value of $X024$ in the previous calculation.

The correction of this error changes the $X024$ integral 
from $-6.0902~(246)$ to $-7.3571~(178)$.
We carefully examined the mapping parts of the other 388 integrals 
and found no error.
The new numerical values of these 388 integrals are consistent with the previous results
as shown in Table~\ref{table:X001} that lists
the results of old evaluation, 
except that the wrong value of $X024$ is removed,
the results of new independent evaluation, and the statistical combination of these two sets of results.


\afterpage{
\input table_setV.tex

}

\section{Residual renormalization terms}

To obtain the physical contribution of Set~V, the standard on-shell 
renormalization prescription has to be followed. 
The $\bm{K}$-operation prescription adopted in the numerical 
evaluation, however, 
is different from that, and thus must be adjusted for the difference. 
The IR subtraction terms introduced by $\bm{R}$- and/or $\bm{I}$-operations 
to make each diagram IR-finite should also be restored.
The IR divergences that reside in these two types of adjustments 
compensate with each other among several diagrams of Set V, leaving finite terms
behind. 
This residual renormalization leads to the expression of the physical contribution 
$ A_1^{(10)}[\mbox{Set V}]$  given by
\begin{eqnarray}
A_1^{(10)}[\mbox{Set V}] &=& \Delta M_{10}
    \nonumber \\
   &+& \Delta M_8 ~ (  - 7\DLB_2 )
   \nonumber \\
   &+& \Delta M_6 ~ \{  - 5\DLB_4 + 20(\DLB_2)^2 \}
   \nonumber \\
      &+& \Delta M_4 ~ \{ - 3\DLB_6 + 24\DLB_4~\DLB_2 - 28(\DLB_2)^3 \}
   \nonumber \\
   &+& \Delta M_4~(2 \Delta dm_4 ~ \Delta L_{2^*} )
   \nonumber \\
    &+& M_2 ~ \{  - \DLB_8 + 8\DLB_6~\DLB_2 - 28\DLB_4~(\DLB_2)^2 
   \nonumber \\
   &   &~~~~~~+ 4  (\DLB_4)^2 + 14(\DLB_2)^4 \}
   \nonumber \\
   &+& M_2~\Delta dm_6 ~( 2\Delta L_{2^*} )
   \nonumber \\
   &+& M_2~\Delta dm_4 ~(  - 16\DLB_2~\Delta L_{2^*} - 2\Delta dm_{2^*}~\Delta L_{2^*} + \Delta L_{4^*}     )
   ~. 
 \label{eq:remainder}
\end{eqnarray}
$\Delta M_n$, $\Delta LB_n$, $\Delta dm_n$
are obtained from the magnetic moment amplitudes, the sum of
vertex and wave-function renormalization constants, and
the mass renormalization constants, respectively, of the $n$th-order diagrams
without a fermion loop.  Here the renormalization condition is the standard on-shell one. 
An asterisk 
($*$) indicates that the quantity be derived
from diagrams having a two-point vertex insertion.
UV divergences of all quantities are removed by the $\bm{K}$-operations,
while IR divergences are removed by the $\bm{R}$- and/or $\bm{I}$-operations.
The overall IR divergences in the vertex and wave-function renormalization constants $L_n$ and
$B_n$, respectively,  cannot be handled by $\bm{R}$- nor $\bm{I}$-operations.
They are, however, canceled when two are summed because of the Ward-Takahashi identity. 
The precise definitions of these symbols are given in 
Refs.~\cite{Aoyama:2005kf,Aoyama:2007bs}.
Their numerical values are listed in Table~\ref{table:residual}, 
where most of the numbers are copied 
from Table~II of Ref.~\cite{Aoyama:2014sxa}.
Now that the eighth-order contribution  has been confirmed
\cite{Laporta:2017okg}, it implies that the residual renormalization
constants used to derive \eqref{eq:our8th} should also be correct. 
The value $\Delta M_8 = 1.738~12~(85)$ in the previous work  is replaced by
a new and more accurate value $\Delta M_8 = 1.738~467~(20)$ derived from the almost exact 
eighth-order result Eq.~(5) of Ref.~\cite{Laporta:2017okg}.  

Among many quantities appearing in \eqref{eq:remainder},  only
three, $\Delta L_{4*}$, $\Delta M_{10}$, and $\DLB_8$,  
have not been  independently checked. The quantity $\Delta L_{4^*}$ is derived
from the fourth-order vertex diagrams with a two-point vertex insertion
and was obtained by carrying out small and quick numerical calculations.
The finite magnetic moment amplitude of 
tenth order $\Delta M_{10}$ 
is obtained by summing up the last columns of 
Table~\ref{table:X001}.
The finite part of the vertex and wave-function renormalization constants of  eighth 
order $\DLB_8$ is also obtained by numerical means.

\input table_residual.tex

\input table_DLB8.tex

To obtain the value of $\DLB_8$, we have to deal with 518  eighth-order
vertex diagrams and 74 self-energy diagrams. The Ward-Takahashi identity 
relates seven vertex diagrams to one self-energy diagram. 
Because  of the  time-reversal symmetry,
the number of self-energy-like diagrams is reduced to 47.  
Extraction of the finite part of the renormalization constants in a form of
Feynman-parametric integral from a self-energy-like diagram
was carried out by using an automatic code generator
similar to {\sc gencode}{\it N}. 
The validity of the code generator has been 
checked for the sixth-order quantity $\DLB_6$, which leads to the 
correct eighth-order contribution \eqref{eq:our8th}.
The divergence structure of the 47 finite integrals of $\DLB_8$ and the 8 integrals of $\DLB_6$ 
can be symbolically expressed as shown by Eq.~(B14) 
of Ref.~\cite{Aoyama:2014sxa}.
The 47 integrals, which are 10-dimensional integrals, 
are then numerically evaluated by VEGAS. 
Their values are listed in Table~\ref{table:DLB8}.

Collecting all these results and substituting numerical 
values into Eq.~\eqref{eq:remainder}, 
we obtain \eqref{eq:SetVbest} as the best estimate for the Set~V contribution to $a_e$.

\section{Conclusion}
 
 We have re-evaluated all 389 integrals representing the tenth-order Set~V diagrams.
 The error in the integral $X024$ was identified and corrected. 
For the other 388 integrals, different sets of integration variables 
were prepared and independent numerical evaluation was performed.
 More statistics have been accumulated which enabled 
us to obtain the more reliable  and accurate numerical results.
 Together with the semi-analytic eighth-order result, the uncertainty of the  QED contribution to
 the electron anomalous magnetic moment ($a_e$) has been reduced
to the same order of uncertainty as that of the hadronic contribution.

\begin{acknowledgments}
We thank Dr.~F.~Jegerlehner for providing us the details of his calculation on the
hadronic contribution to $a_e$  prior to its publication.  We also thank Dr.~B.~Taylor for 
notifying us of  the new CODATA adjustment of fundamental physical constants.
We thank Dr.~A.~Kataev, Dr.~M.~Hoferichter, and Dr.~D.~Nomura for communication.
T.~K.'s work is supported in part by the U. S. National Science Foundation
under Grant No. NSF-PHY-1316222.
M.~N.'s work is supported in part by the JSPS Grant-in-Aid for Scientific Research (C)16K05338.
Numerical calculations were conducted on the supercomputing systems
on RICC, HOKUSAI-GreatWave, and HOKUSAI-BigWaterfall at RIKEN.
\end{acknowledgments}


\appendix

\section{Updates for  the muon $g-2$}
\label{sec:appendix_amu}

The world average of the measurements of  the muon anomalous magnetic moment ($a_\mu$) is 
dominated by the BNL experiments and is given as \cite{Bennett:2006fi, Mohr:2015ccw}
\begin{equation}
a_\mu(\text{expt.}) = 116~592~089~(63) \times 10^{-11} ~[0.5 \text{ppm}]~.
\end{equation}
A new experiment is currently being prepared at Fermilab using the storage ring that was 
used for the previous BNL measurements and was moved from BNL\cite{Grange:2015fou}.  The first result of the
Fermilab E989 experiment is expected to be announced in Spring 2018 \cite{Chapelain:2017syu}.
Another new experiment is being prepared at J-PARC\cite{Iinuma:2011zz}.  Cold muon beam
and a storage ring with 66 cm diameter are used for this E34 experiment \cite{Otani:2015jra}.
Two measurements with completely different apparatus  will provide us 
far deeper insight on the physics of muon $g\!-\!2$. 

The QED contribution to 
$a_\mu$ is affected
by the improved values of the mass-independent 
eighth- and tenth-order terms and by the new values of the fine-structure constant.  
Two independent determinations  of  $\alpha$ lead to two values of the QED contribution of $a_\mu$.
By using the $\alpha$ from the Rb-atom experiment \eqref{eq:alinvRb2016} or that from the electron $g\!-\!2$ 
\eqref{eq:alinv_ae}, the QED contribution of $a_\mu$ is given as
\begin{eqnarray}
a_\mu(\text{QED},\alpha(\text{Rb}))& =&  1~165~847~189.71~ (7) (17) (6) (72) \times 10^{-12}, 
\label{eq:amu_QED_Rb}
\\
&& \text{or} \nonumber \\
a_\mu(\text{QED},\alpha(a_e))         &=& 1~165~847~188.41~(7) (17) (6) (28) \times 10^{-12},
\label{eq:amu_QED_ae}
\end{eqnarray}
respectively, where the  assigned uncertainties are due to the lepton-mass ratios, 
the numerical evaluation of the
eighth-order QED term, 
the numerical evaluation of the tenth-order QED term, 
and the fine-structure constant from left to right.
By now the entire eighth-order QED term has been cross-checked by two or more independent
calculations.
The mass-dependent eighth-order QED terms involving the electron or tau-lepton 
have been calculated by numerical means \cite{Kinoshita:2004wi, amu10:PRL} and the results are confirmed 
by analytic calculation \cite{Kurz:2016bau,Kurz:2015bia,Kurz:2013exa}. 
The QED predictions \eqref{eq:amu_QED_Rb} and \eqref{eq:amu_QED_ae} are 
changed by $+0.20 \times 10^{-12}$ and $-0.05 \times 10^{-12}$, respectively, 
from those given in Ref. \cite{amu10:PRL}, 
which does not affect comparison of theory and measurement of $a_\mu$.

To compare theory of $a_\mu$ to measurement, the hadronic and electroweak contributions 
must be added to the dominant QED contribution \eqref{eq:amu_QED_Rb} or \eqref{eq:amu_QED_ae}.
The hadronic vacuum-polarization (HVP) contribution has been calculated by three
groups based on newest measurements of the hadronic cross section. 
The LO-HVP contribution
is found  in Refs. \cite{Jegerlehner:2017zsb,Teubner:2017phipsi,Davier:2017zfy}.
Recently a remarkable progress has been achieved in lattice calculation of the LO-HVP
contribution \cite{Chakraborty:2016mwy, Blum:2015you}, although it is not yet competitive in precision
to the dispersion calculation based on measured hadronic cross sections.
Reevaluation of the NLO-HVP contribution by means of the dispersion integral is given in 
Refs.~\cite{Jegerlehner:2017zsb, Teubner:2017phipsi}.
The NNLO-HVP contribution is given in Ref.~\cite{Kurz:2014wya}.
All the HVP contributions are summarized as 
\begin{eqnarray}
a_\mu(\text{HVP, LO}) & = & \left \{  
                                           \begin{array}{ll}
                                               689.46~( 3.25 ) \times 10^{-10}  & ~~~~~\text{\cite{Jegerlehner:2017zsb}}  \\
                                               692.23~(2.54)   \times 10^{-10}  & ~~~~~\text{\cite{Teubner:2017phipsi}} \\
                                               693.1~(3.4) \times 10^{-10}        & ~~~~~\text{\cite{Davier:2017zfy}}
                                          \end{array}
                                            \right .
                                            ,
\label{eq:amu_HVP_LO}
                                  \\
a_\mu(\text{HVP, NLO}) & = & \left\{
                                            \begin{array}{ll}
                                                 -9.927~(0.067) \times 10^{-10} & ~~~~~\text{\cite{Jegerlehner:2017zsb}}  \\
                                                 -9.83~(0.04) \times 10^{-10} & ~~~~~\text{\cite{Teubner:2017phipsi}}
                                            \end{array}
                                                   \right .
                                                   ,
\label{eq:amu_HVP_NLO}
                                   \\         
a_\mu(\text{HVP, NNLO}) &=& \left . ~~~~~
                                           \begin{array}{ll} 
                                                 1.24~(0.01) \times 10^{-10}  & ~~~~~~~\text{\cite{Kurz:2014wya}} 
                                           \end{array}  
                                           \right .
                                           .    
\label{eq:amu_HVP_NNLO}
\end{eqnarray}

The largest uncertainty of the theory of $a_\mu$ comes from the hadronic light-by-light (HLbL) contribution.
Various hadron models have been used to compute it. The widely accepted value of the LO-HLbL term is
$10.5 (2.6) \times 10^{-10}$ \cite{Prades:2009tw} or
$11.6 (4.0)\times 10^{-10}$ \cite{Jegerlehner:2009ry}. Both cover almost all model-dependent results \cite{Bijnens:1995cc,Bijnens:1995xf,Hayakawa:1995ps,Hayakawa:1996ki,Hayakawa:1997rq,Melnikov:2003xd,Bijnens:2007pz,Nyffeler:2009tw}.
Recently, the axial meson
contribution to the HLbL has been revised \cite{Pauk:2014rta}, which  makes the LO-HLbL contribution smaller.  
Other revisions have been made for
the tensor-exchange contribution \cite{Pauk:2014rta} and for the $\pi^0$ exchange contribution based on the
lattice calculation
\cite{Gerardin:2016cqj}.  By collecting these modifications,  the LO-HLbL becomes \cite{Jegerlehner:2017lbd}
\begin{equation}
a_\mu(\text{HLbL, LO}) = 10.0 ~(2.9) \times 10^{-10}.
\label{eq:amu_HLbL_LO}
\end{equation}
The NLO-HLbL term was estimated to be \cite{Colangelo:2014qya}  
\begin{equation} 
a_\mu(\text{HLbL, NLO}) = 0.3~(0.2) \times 10^{-10}.
\label{eq:amu_HLbL_NLO}
\end{equation}

In last few years, lattice QCD calculation of the HLbL contribution has become feasible \cite{Green:2015sra, Asmussen:2016lse, Blum:2016lnc, Blum:2017cer,Gerardin:2017ryf}.
Though the systematic uncertainty has not yet been determined, Ref.~\cite{Blum:2016lnc} gives the lattice QCD result  as
\begin{equation}
  a_\mu(\text{HLbL, LO}) = 5.35~(1.35) \times 10^{-10}~,
\label{eq:amu_HLbL_lattice}
\end{equation}
where the uncertainty is due to statistics only.

The data-driven dispersion approach to HLbL similar to but far more complicated than that to HVP has been formulated \cite{Colangelo:2014dfa,Pauk:2014rfa}. 
It is a very promissing method though it requires long and difficult theoretical works\cite{Colangelo:2015ama,Colangelo:2014pva}. The contribution from the two-$\pi$ intermediate states has been recently determined\cite{Colangelo:2017fiz}, which is more accurate
than that determined using the hadronic models.

For the electroweak contribution, up to two-loop calculations are known \cite{Fujikawa:1972fe,Czarnecki:2002nt}.
The two-loop weak correction was reevaluated using the latest measured mass of the Higgs boson \cite{Gnendiger:2013pva} and the total contribution becomes 
\begin{equation}
a_\mu(\text{Weak}) = 15.36~(10) \times 10^{-10}.
\label{eq:amu_weak}
\end{equation}

The theoretical prediction of $a_\mu$ are obtained by summing up the QED, hadronic, and weak contributions.
We use the model calculation \eqref{eq:amu_HLbL_LO} for the LO-HLbL contribution.
The three values of the LO-HVP estimates listed in Eq.~\eqref{eq:amu_HVP_LO} lead  the theoretical prediction of $a_\mu$ as 
\begin{equation}
a_\mu(\text{theory})  =  \left \{  
                                     \begin{array}{ll}
                                                116~591~783~(51)   \times 10^{-11}  & ~~~~~\text{\cite{Jegerlehner:2017zsb} }  \\
                                                116~591~812~(47)   \times 10^{-11}  & ~~~~~\text{\cite{Teubner:2017phipsi}} \\
                                                116~591~820~(52)    \times 10^{-11}        & ~~~~~\text{\cite{Davier:2017zfy}}
                                      \end{array}         
                               \right .
                               .
\end{equation}
The difference between  measurement and theory ranges from $269 \sim 306\times 10^{-11}$ corresponding
to  $3.3 \sim 3.8~ \sigma$ discrepancy.


\bibliographystyle{apsrev}
\bibliography{b}

\end{document}

%% file: table_setV.tex
\begingroup
\squeezetable
\setlength\LTleft{0pt}
\setlength\LTright{0pt}
\setlength\LTcapwidth{\textwidth}
\begin{longtable*}{l@{\hskip-9em}l@{\hskip-5em}d@{\hskip-1em}d@{\hskip-1em}d@{\hskip-1em}d}
\caption*{%
\small TABLE \ref{table:X001}:
VEGAS integration results of $X001$--$X389$ of the tenth-order Set V diagrams.
The first and second columns show the diagram name and 
its representation in terms of photon indices, respectively.
The third and fourth columns list the VEGAS integration results 
used in Ref. \cite{Aoyama:2014sxa}
and the new VEGAS integration result of this work, respectively.
An uncertainty of the integral corresponds to 
a  $1.65 \sigma$, which is 90\% confidence level,  
determined by VEGAS 
assuming  gaussian distribution of statistical ensembles for the integral.  
The difference of two results is listed in the fifth column.
The weighted average of the third and fourth columns is listed in the last column.
The old result of $X024$ from Ref. \cite{Aoyama:2014sxa} is incorrect and 
is removed from the third column.
The integral $X024$ is thus not averaged over the third and fourth columns.
\label{table:X001}
} \\
\hline \hline
\multicolumn{1}{p{0.12\textwidth}}{Diagram $\mathcal{G}$} &
\multicolumn{1}{p{0.12\textwidth}}{Expression} &
\multicolumn{1}{p{0.18\textwidth}}{\hspace{3em}\begin{tabular}{l} Value (Error) \\ 
                                                      in Ref. \cite{Aoyama:2014sxa}
                                              \end{tabular}
                                   } &
\multicolumn{1}{p{0.18\textwidth}}{\hspace{3em}
                                   \begin{tabular}{l} Value (Error) \\ 
                                                      in this work 
                                   \end{tabular}
                                   } &
\multicolumn{1}{p{0.18\textwidth}}{\hspace{4em}Difference} &
\multicolumn{1}{p{0.18\textwidth}}{\hspace{3em}
                                   \begin{tabular}{l} Weighted \\ 
                                                      average 
                                   \end{tabular}
                                   } 
\\
\hline
\endfirsthead
\caption*{%
\small TABLE \ref{table:X001}(continued):
VEGAS integration results of $X001$--$X389$ of the tenth-order Set V diagrams.
} \\
\hline \hline
\multicolumn{1}{p{0.12\textwidth}}{Diagram $\mathcal{G}$} &
\multicolumn{1}{p{0.12\textwidth}}{Expression} &
\multicolumn{1}{p{0.18\textwidth}}{\hspace{3em}
                                   \begin{tabular}{l} Value (Error) \\ 
                                                      in Ref. \cite{Aoyama:2014sxa}
                                   \end{tabular}
                                   } &
\multicolumn{1}{p{0.18\textwidth}}{\hspace{3em}\begin{tabular}{l} Value (Error) \\ 
                                                      in this work 
                                   \end{tabular}
                                   } &
\multicolumn{1}{p{0.18\textwidth}}{\hspace{4em}Difference} &
\multicolumn{1}{p{0.18\textwidth}}{\hspace{3em}\begin{tabular}{l} Weighted \\ 
                                                      average 
                                   \end{tabular}
                                   } 
\\
\hline
\endhead
\hline
\endfoot
\hline \hline
\endlastfoot
$X001$~& $abacbdcede$ &   -0.1724~( 92) &   -0.1591~( 35) &   -0.0133 &   -0.1608~( 33) \\
$X002$~& $abaccddebe$ &   -5.9958~(333) &   -5.9488~(176) &   -0.0471 &   -5.9591~(156) \\
$X003$~& $abacdbcede$ &   -0.1057~( 52) &   -0.1048~( 18) &   -0.0009 &   -0.1049~( 17) \\
$X004$~& $abacdcdebe$ &    5.1027~(339) &    5.1019~(182) &    0.0007 &    5.1021~(160) \\
$X005$~& $abacddbece$ &    1.1112~(168) &    1.0973~(212) &    0.0138 &    1.1058~(131) \\
$X006$~& $abacddcebe$ &   -5.2908~(245) &   -5.2942~(215) &    0.0033 &   -5.2927~(161) \\
$X007$~& $abbcadceed$ &   -3.4592~(254) &   -3.4319~(217) &   -0.0273 &   -3.4434~(165) \\
$X008$~& $abbccddeea$ &  -16.5070~(289) &  -16.4999~(201) &   -0.0071 &  -16.5022~(165) \\
$X009$~& $abbcdaceed$ &   -3.1069~( 71) &   -3.1336~(174) &    0.0267 &   -3.1107~( 65) \\
$X010$~& $abbcdcdeea$ &   11.2644~(342) &   11.2817~(178) &   -0.0173 &   11.2780~(158) \\
$X011$~& $abbcddaeec$ &    6.0467~(338) &    6.0445~(183) &    0.0022 &    6.0450~(161) \\
$X012$~& $abbcddceea$ &   -9.3328~(267) &   -9.3587~(211) &    0.0259 &   -9.3488~(166) \\
$X013$~& $abcabdecde$ &   -1.3710~( 31) &   -1.3759~(  9) &    0.0049 &   -1.3755~(  9) \\
$X014$~& $abcacdedbe$ &    0.8727~( 42) &    0.8789~( 15) &   -0.0062 &    0.8782~( 14) \\
$X015$~& $abcadbecde$ &    2.1090~(  8) &    2.1107~(  4) &   -0.0017 &    2.1103~(  4) \\
$X016$~& $abcadcedbe$ &   -0.9591~(  7) &   -0.9588~(  3) &   -0.0003 &   -0.9588~(  3) \\
$X017$~& $abcaddebce$ &    0.5146~( 13) &    0.5162~( 20) &   -0.0016 &    0.5151~( 11) \\
$X018$~& $abcaddecbe$ &    0.0309~( 13) &    0.0323~( 21) &   -0.0014 &    0.0313~( 11) \\
$X019$~& $abcbadeced$ &    1.2965~( 48) &    1.3028~( 15) &   -0.0063 &    1.3022~( 14) \\
$X020$~& $abcbcdedea$ &   -8.1900~(318) &   -8.1534~(187) &   -0.0366 &   -8.1628~(161) \\
$X021$~& $abcbdaeced$ &   -0.2948~( 15) &   -0.2954~(  9) &    0.0006 &   -0.2952~(  8) \\
$X022$~& $abcbdcedea$ &    0.8892~(226) &    0.8839~(211) &    0.0053 &    0.8864~(154) \\
$X023$~& $abcbddeaec$ &    0.4485~( 55) &    0.4543~(103) &   -0.0058 &    0.4498~( 49) \\
$X024$~& $abcbddecea$ &    -           &   -7.3481~(139) &    -      &   -7.3481~(139) \\
$X025$~& $abccadeebd$ &   -0.7482~(194) &   -0.7585~(219) &    0.0103 &   -0.7528~(145) \\
$X026$~& $abccbdeeda$ &   -7.8258~(277) &   -7.8213~(210) &   -0.0045 &   -7.8230~(167) \\
$X027$~& $abccdaeebd$ &   -2.3260~( 54) &   -2.3185~( 68) &   -0.0075 &   -2.3231~( 42) \\
$X028$~& $abccdbeeda$ &    4.5663~(342) &    4.5459~(177) &    0.0204 &    4.5502~(157) \\
$X029$~& $abccddeeab$ &    6.9002~(233) &    6.9066~(183) &   -0.0064 &    6.9042~(144) \\
$X030$~& $abccddeeba$ &  -12.6225~(342) &  -12.6385~(193) &    0.0160 &  -12.6346~(168) \\
$X031$~& $abcdaebcde$ &    2.3000~( 14) &    2.3011~(  8) &   -0.0011 &    2.3009~(  6) \\
$X032$~& $abcdaecdbe$ &   -0.2414~(  6) &   -0.2422~(  4) &    0.0008 &   -0.2420~(  3) \\
$X033$~& $abcdaedbce$ &   -1.3806~(  7) &   -1.3809~(  4) &    0.0003 &   -1.3808~(  3) \\
$X034$~& $abcdaedcbe$ &    1.2585~(  9) &    1.2599~(  5) &   -0.0014 &    1.2595~(  4) \\
$X035$~& $abcdbeaced$ &   -0.5899~(  3) &   -0.5899~(  2) &   -0.0000 &   -0.5899~(  1) \\
$X036$~& $abcdbecdea$ &    0.2318~( 11) &    0.2327~( 22) &   -0.0009 &    0.2320~( 10) \\
$X037$~& $abcdbedaec$ &   -0.7407~(  5) &   -0.7410~(  2) &    0.0003 &   -0.7409~(  2) \\
$X038$~& $abcdbedcea$ &   -0.2927~( 14) &   -0.2919~( 21) &   -0.0008 &   -0.2924~( 11) \\
$X039$~& $abcdceaebd$ &    0.3292~( 12) &    0.3307~(  9) &   -0.0015 &    0.3301~(  7) \\
$X040$~& $abcdcebeda$ &    1.3397~( 50) &    1.3434~( 66) &   -0.0037 &    1.3411~( 40) \\
$X041$~& $abcdcedeab$ &    3.1076~( 94) &    3.1219~(160) &   -0.0143 &    3.1113~( 81) \\
$X042$~& $abcdcedeba$ &   -4.1353~(192) &   -4.1235~(218) &   -0.0119 &   -4.1301~(144) \\
$X043$~& $abcddeeabc$ &   -2.9620~( 29) &   -2.9633~( 53) &    0.0013 &   -2.9623~( 25) \\
$X044$~& $abcddeebca$ &    4.4121~(281) &    4.3654~(160) &    0.0467 &    4.3769~(139) \\
$X045$~& $abcddeecab$ &    3.4331~(212) &    3.4338~(206) &   -0.0007 &    3.4335~(148) \\
$X046$~& $abcddeecba$ &   -7.7564~(339) &   -7.7945~(187) &    0.0382 &   -7.7856~(163) \\
$X047$~& $abcdeabcde$ &   -4.4496~( 40) &   -4.4602~( 11) &    0.0106 &   -4.4594~( 11) \\
$X048$~& $abcdeacdbe$ &   -0.8061~(  8) &   -0.8058~(  4) &   -0.0003 &   -0.8058~(  3) \\
$X049$~& $abcdeadbce$ &   -0.0278~(  7) &   -0.0280~(  3) &    0.0003 &   -0.0280~(  3) \\
$X050$~& $abcdeadcbe$ &   -1.2213~(  9) &   -1.2213~(  5) &   -0.0000 &   -1.2213~(  4) \\
$X051$~& $abcdebaced$ &   -0.1776~(  6) &   -0.1774~(  4) &   -0.0001 &   -0.1775~(  3) \\
$X052$~& $abcdebcdea$ &    1.0293~( 17) &    1.0279~( 30) &    0.0014 &    1.0289~( 15) \\
$X053$~& $abcdebdaec$ &    0.3699~(  4) &    0.3702~(  2) &   -0.0002 &    0.3701~(  2) \\
$X054$~& $abcdebdcea$ &   -0.5174~( 11) &   -0.5196~( 20) &    0.0023 &   -0.5179~(  9) \\
$X055$~& $abcdecaebd$ &   -0.3673~(  4) &   -0.3679~(  2) &    0.0006 &   -0.3677~(  1) \\
$X056$~& $abcdecbeda$ &   -0.2650~( 27) &   -0.2608~( 42) &   -0.0042 &   -0.2637~( 23) \\
$X057$~& $abcdecdeab$ &    2.7370~( 31) &    2.7385~( 58) &   -0.0015 &    2.7373~( 27) \\
$X058$~& $abcdecdeba$ &   -5.2510~( 70) &   -5.2249~(140) &   -0.0261 &   -5.2457~( 63) \\
$X059$~& $abcdedeabc$ &    2.1866~( 28) &    2.1758~( 50) &    0.0108 &    2.1841~( 24) \\
$X060$~& $abcdedebca$ &   -3.2089~(188) &   -3.1792~(216) &   -0.0297 &   -3.1962~(142) \\
$X061$~& $abcdedecab$ &   -3.7724~(137) &   -3.7874~(216) &    0.0149 &   -3.7767~(116) \\
$X062$~& $abcdedecba$ &    5.9174~(262) &    5.8861~(219) &    0.0313 &    5.8990~(168) \\
$X063$~& $abcdeeabcd$ &    3.4295~( 14) &    3.4297~( 26) &   -0.0002 &    3.4296~( 12) \\
$X064$~& $abcdeeacbd$ &   -0.2772~(  8) &   -0.2779~( 14) &    0.0008 &   -0.2774~(  7) \\
$X065$~& $abcdeebadc$ &    0.1551~( 13) &    0.1580~( 21) &   -0.0029 &    0.1559~( 11) \\
$X066$~& $abcdeebcda$ &   -3.6145~( 45) &   -3.6177~( 81) &    0.0033 &   -3.6152~( 39) \\
$X067$~& $abcdeecdab$ &   -1.6761~( 85) &   -1.6853~(168) &    0.0092 &   -1.6780~( 76) \\
$X068$~& $abcdeecdba$ &    2.7855~(217) &    2.7540~(205) &    0.0315 &    2.7689~(149) \\
$X069$~& $abcdeedabc$ &   -1.2627~( 31) &   -1.2690~( 45) &    0.0063 &   -1.2647~( 25) \\
$X070$~& $abcdeedbca$ &    3.2149~(144) &    3.2001~(212) &    0.0148 &    3.2102~(119) \\
$X071$~& $abcdeedcab$ &    3.7025~( 96) &    3.6943~(187) &    0.0083 &    3.7008~( 85) \\
$X072$~& $abcdeedcba$ &   -5.5704~(208) &   -5.5658~(209) &   -0.0047 &   -5.5681~(147) \\
$X073$~& $abacbdceed$ &    3.4114~(254) &    3.3929~(212) &    0.0184 &    3.4005~(162) \\
$X074$~& $abacbddece$ &    4.4104~(251) &    4.3889~(212) &    0.0215 &    4.3978~(162) \\
$X075$~& $abacbddeec$ &   -8.1138~(340) &   -8.0608~(195) &   -0.0531 &   -8.0739~(169) \\
$X076$~& $abacbdecde$ &   -5.3405~( 74) &   -5.3407~( 31) &    0.0003 &   -5.3407~( 29) \\
$X077$~& $abacbdeced$ &    3.5459~( 86) &    3.5604~( 51) &   -0.0146 &    3.5567~( 43) \\
$X078$~& $abacbdedce$ &    1.1666~( 80) &    1.1778~( 48) &   -0.0112 &    1.1748~( 41) \\
$X079$~& $abacbdedec$ &    5.3956~(305) &    5.4128~(205) &   -0.0173 &    5.4075~(170) \\
$X080$~& $abacbdeecd$ &    0.4597~(257) &    0.4648~(217) &   -0.0051 &    0.4627~(166) \\
$X081$~& $abacbdeedc$ &   -5.6566~(248) &   -5.6298~(217) &   -0.0268 &   -5.6414~(163) \\
$X082$~& $abaccdbeed$ &   -8.5156~(348) &   -8.4810~(195) &   -0.0345 &   -8.4893~(170) \\
$X083$~& $abaccddeeb$ &   18.7464~(346) &   18.7522~(207) &   -0.0057 &   18.7507~(177) \\
$X084$~& $abaccdebde$ &    8.9888~(129) &    8.9968~(209) &   -0.0080 &    8.9911~(110) \\
$X085$~& $abaccdebed$ &   -2.2833~(197) &   -2.2933~(213) &    0.0100 &   -2.2879~(144) \\
$X086$~& $abaccdedbe$ &    0.5180~(223) &    0.5162~(218) &    0.0018 &    0.5171~(155) \\
$X087$~& $abaccdedeb$ &  -16.5849~(349) &  -16.5942~(173) &    0.0093 &  -16.5923~(155) \\
$X088$~& $abaccdeebd$ &   -5.2606~(340) &   -5.2320~(197) &   -0.0286 &   -5.2392~(171) \\
$X089$~& $abaccdeedb$ &   12.6789~(341) &   12.6723~(194) &    0.0066 &   12.6739~(169) \\
$X090$~& $abacdbceed$ &    1.5206~(130) &    1.5285~(211) &   -0.0079 &    1.5228~(111) \\
$X091$~& $abacdbdece$ &   -1.6355~( 97) &   -1.6320~( 58) &   -0.0035 &   -1.6330~( 50) \\
$X092$~& $abacdbdeec$ &    2.1303~(218) &    2.1083~(201) &    0.0220 &    2.1184~(147) \\
$X093$~& $abacdbecde$ &   -1.7594~( 42) &   -1.7538~( 16) &   -0.0056 &   -1.7545~( 15) \\
$X094$~& $abacdbeced$ &   -1.0419~( 67) &   -1.0406~( 20) &   -0.0014 &   -1.0407~( 19) \\
$X095$~& $abacdbedce$ &    0.5838~( 35) &    0.5875~( 11) &   -0.0037 &    0.5872~( 11) \\
$X096$~& $abacdbedec$ &    1.3458~( 73) &    1.3495~( 22) &   -0.0037 &    1.3492~( 21) \\
$X097$~& $abacdbeecd$ &    5.0319~( 89) &    5.0183~(195) &    0.0136 &    5.0296~( 81) \\
$X098$~& $abacdbeedc$ &   -1.9806~(183) &   -2.0218~(215) &    0.0411 &   -1.9979~(139) \\
$X099$~& $abacdcbeed$ &    3.0771~(187) &    3.0553~(218) &    0.0218 &    3.0678~(142) \\
$X100$~& $abacdcdeeb$ &  -15.2919~(331) &  -15.2360~(203) &   -0.0559 &  -15.2513~(173) \\
$X101$~& $abacdcebde$ &   -0.2462~( 64) &   -0.2397~( 26) &   -0.0065 &   -0.2406~( 24) \\
$X102$~& $abacdcebed$ &   -1.2883~( 75) &   -1.2953~( 34) &    0.0070 &   -1.2941~( 31) \\
$X103$~& $abacdcedbe$ &    0.9424~( 74) &    0.9482~( 21) &   -0.0057 &    0.9477~( 20) \\
$X104$~& $abacdcedeb$ &    6.4131~(298) &    6.3706~(217) &    0.0426 &    6.3853~(175) \\
$X105$~& $abacdceebd$ &    3.0503~(215) &    3.0491~(216) &    0.0012 &    3.0497~(152) \\
$X106$~& $abacdceedb$ &  -11.5662~(344) &  -11.5384~(201) &   -0.0277 &  -11.5455~(174) \\
$X107$~& $abacddbeec$ &   -4.6573~(345) &   -4.6265~(193) &   -0.0308 &   -4.6338~(168) \\
$X108$~& $abacddceeb$ &   12.9775~(341) &   12.9927~(193) &   -0.0152 &   12.9890~(168) \\
$X109$~& $abacddebce$ &   -0.0860~( 85) &   -0.0841~(182) &   -0.0019 &   -0.0857~( 77) \\
$X110$~& $abacddebec$ &    1.9248~(204) &    1.9014~(204) &    0.0234 &    1.9131~(144) \\
$X111$~& $abacddecbe$ &    3.3578~(132) &    3.3641~(190) &   -0.0062 &    3.3599~(108) \\
$X112$~& $abacddeceb$ &  -11.8998~(332) &  -11.8990~(208) &   -0.0008 &  -11.8992~(176) \\
$X113$~& $abacddeebc$ &   -4.3847~(322) &   -4.4412~(176) &    0.0565 &   -4.4282~(155) \\
$X114$~& $abacddeecb$ &   11.0641~(343) &   11.0287~(196) &    0.0355 &   11.0374~(170) \\
$X115$~& $abacdebcde$ &   -0.5974~( 52) &   -0.6028~( 20) &    0.0054 &   -0.6020~( 19) \\
$X116$~& $abacdebced$ &    1.8362~( 28) &    1.8400~( 11) &   -0.0038 &    1.8394~( 10) \\
$X117$~& $abacdebdce$ &    0.3292~( 27) &    0.3309~( 10) &   -0.0016 &    0.3307~(  9) \\
$X118$~& $abacdebdec$ &   -3.2721~( 55) &   -3.2764~( 16) &    0.0043 &   -3.2761~( 16) \\
$X119$~& $abacdebecd$ &   -0.0751~( 53) &   -0.0716~( 18) &   -0.0035 &   -0.0720~( 17) \\
$X120$~& $abacdebedc$ &    1.8769~( 72) &    1.8847~( 22) &   -0.0078 &    1.8840~( 21) \\
$X121$~& $abacdecbde$ &   -0.8549~( 43) &   -0.8511~( 14) &   -0.0039 &   -0.8515~( 13) \\
$X122$~& $abacdecbed$ &   -0.7337~( 42) &   -0.7346~( 15) &    0.0008 &   -0.7345~( 14) \\
$X123$~& $abacdecdbe$ &   -3.3559~( 67) &   -3.3564~( 21) &    0.0004 &   -3.3563~( 20) \\
$X124$~& $abacdecdeb$ &   11.5746~(106) &   11.5778~(204) &   -0.0033 &   11.5752~( 94) \\
$X125$~& $abacdecebd$ &    0.8677~( 64) &    0.8710~( 19) &   -0.0033 &    0.8707~( 18) \\
$X126$~& $abacdecedb$ &   -1.5696~(162) &   -1.5809~(199) &    0.0113 &   -1.5741~(125) \\
$X127$~& $abacdedbce$ &    1.1412~( 46) &    1.1495~( 17) &   -0.0083 &    1.1484~( 16) \\
$X128$~& $abacdedbec$ &    0.6493~( 59) &    0.6521~( 17) &   -0.0027 &    0.6518~( 16) \\
$X129$~& $abacdedcbe$ &    1.4833~( 70) &    1.4890~( 21) &   -0.0057 &    1.4885~( 20) \\
$X130$~& $abacdedceb$ &   -1.5696~(180) &   -1.5797~(205) &    0.0102 &   -1.5740~(135) \\
$X131$~& $abacdedebc$ &    3.1060~(287) &    3.0832~(219) &    0.0228 &    3.0916~(174) \\
$X132$~& $abacdedecb$ &   -8.8300~(337) &   -8.8562~(198) &    0.0262 &   -8.8495~(170) \\
$X133$~& $abacdeebcd$ &    2.7263~( 88) &    2.7345~(177) &   -0.0082 &    2.7279~( 79) \\
$X134$~& $abacdeebdc$ &   -0.6712~(123) &   -0.6569~(198) &   -0.0143 &   -0.6672~(104) \\
$X135$~& $abacdeecbd$ &    0.9256~(153) &    0.9201~(207) &    0.0054 &    0.9236~(123) \\
$X136$~& $abacdeecdb$ &   -7.5256~(305) &   -7.5147~(205) &   -0.0110 &   -7.5181~(170) \\
$X137$~& $abacdeedbc$ &   -2.3541~(233) &   -2.3413~(209) &   -0.0128 &   -2.3470~(156) \\
$X138$~& $abacdeedcb$ &   10.1610~(284) &   10.1624~(215) &   -0.0014 &   10.1619~(171) \\
$X139$~& $abbcaddeec$ &   14.8650~(348) &   14.8877~(203) &   -0.0227 &   14.8819~(176) \\
$X140$~& $abbcadeced$ &   -2.7901~(206) &   -2.8044~(207) &    0.0143 &   -2.7972~(146) \\
$X141$~& $abbcadedec$ &  -12.5567~(350) &  -12.4879~(207) &   -0.0688 &  -12.5057~(178) \\
$X142$~& $abbcadeecd$ &   -1.5767~(341) &   -1.5679~(202) &   -0.0088 &   -1.5702~(173) \\
$X143$~& $abbcadeedc$ &   10.3225~(341) &   10.3377~(195) &   -0.0152 &   10.3339~(169) \\
$X144$~& $abbccdedea$ &   23.7239~(368) &   23.6713~(189) &    0.0526 &   23.6823~(168) \\
$X145$~& $abbccdeeda$ &  -18.6212~(349) &  -18.6295~(166) &    0.0083 &  -18.6279~(150) \\
$X146$~& $abbcdadeec$ &   -2.2990~(335) &   -2.2458~(202) &   -0.0532 &   -2.2600~(173) \\
$X147$~& $abbcdaeced$ &    1.1243~( 55) &    1.1316~(101) &   -0.0074 &    1.1259~( 48) \\
$X148$~& $abbcdaedec$ &   -1.4150~( 76) &   -1.4002~(127) &   -0.0148 &   -1.4111~( 65) \\
$X149$~& $abbcdaeecd$ &   -8.3898~(139) &   -8.3948~(200) &    0.0050 &   -8.3914~(114) \\
$X150$~& $abbcdaeedc$ &    2.8758~(260) &    2.9171~(200) &   -0.0413 &    2.9017~(158) \\
$X151$~& $abbcdcedea$ &  -10.9362~(344) &  -10.9329~(199) &   -0.0033 &  -10.9337~(172) \\
$X152$~& $abbcdceeda$ &   14.6793~(345) &   14.6367~(201) &    0.0426 &   14.6475~(173) \\
$X153$~& $abbcddecea$ &   14.8936~(343) &   14.8523~(199) &    0.0414 &   14.8627~(172) \\
$X154$~& $abbcddeeca$ &  -20.6285~(342) &  -20.5999~(200) &   -0.0286 &  -20.6072~(173) \\
$X155$~& $abbcdeadec$ &    5.0341~( 46) &    5.0371~( 75) &   -0.0030 &    5.0349~( 39) \\
$X156$~& $abbcdeaedc$ &   -0.8277~( 69) &   -0.8285~(130) &    0.0008 &   -0.8279~( 60) \\
$X157$~& $abbcdecdea$ &  -11.8490~(252) &  -11.8884~(205) &    0.0394 &  -11.8727~(159) \\
$X158$~& $abbcdeceda$ &    0.4607~(329) &    0.4106~(206) &    0.0502 &    0.4247~(174) \\
$X159$~& $abbcdedcea$ &    0.4435~(351) &    0.4270~(173) &    0.0165 &    0.4302~(155) \\
$X160$~& $abbcdedeca$ &   14.0724~(349) &   14.0370~(197) &    0.0354 &   14.0455~(171) \\
$X161$~& $abbcdeecda$ &    7.8073~(342) &    7.7941~(198) &    0.0131 &    7.7974~(171) \\
$X162$~& $abbcdeedca$ &  -12.8293~(339) &  -12.8564~(195) &    0.0271 &  -12.8496~(169) \\
$X163$~& $abcabdceed$ &    6.8168~(202) &    6.8111~(214) &    0.0057 &    6.8141~(147) \\
$X164$~& $abcabddeec$ &  -12.8880~(208) &  -12.8941~(180) &    0.0061 &  -12.8915~(136) \\
$X165$~& $abcabdeced$ &   -2.1661~( 76) &   -2.1641~( 22) &   -0.0020 &   -2.1643~( 21) \\
$X166$~& $abcabdedce$ &   -2.3081~( 70) &   -2.3088~( 25) &    0.0008 &   -2.3087~( 23) \\
$X167$~& $abcabdedec$ &   12.1361~(150) &   12.1348~(196) &    0.0013 &   12.1356~(119) \\
$X168$~& $abcabdeecd$ &    3.4447~(120) &    3.4443~(195) &    0.0004 &    3.4446~(102) \\
$X169$~& $abcabdeedc$ &   -6.9379~(108) &   -6.9384~(193) &    0.0005 &   -6.9380~( 94) \\
$X170$~& $abcacdbeed$ &    0.2635~(288) &    0.2420~(183) &    0.0215 &    0.2482~(154) \\
$X171$~& $abcacddeeb$ &   -2.5229~(313) &   -2.5628~(194) &    0.0399 &   -2.5517~(165) \\
$X172$~& $abcacdebed$ &    1.5601~( 76) &    1.5697~( 32) &   -0.0096 &    1.5683~( 30) \\
$X173$~& $abcacdedeb$ &    0.0193~(298) &   -0.0209~(215) &    0.0401 &   -0.0071~(174) \\
$X174$~& $abcacdeebd$ &    1.7158~(191) &    1.7123~(203) &    0.0035 &    1.7142~(139) \\
$X175$~& $abcacdeedb$ &   -1.8253~(175) &   -1.8346~(206) &    0.0093 &   -1.8292~(133) \\
$X176$~& $abcadbceed$ &    0.7450~( 35) &    0.7430~( 64) &    0.0021 &    0.7446~( 30) \\
$X177$~& $abcadbdeec$ &    0.0079~( 81) &    0.0411~(196) &   -0.0332 &    0.0127~( 74) \\
$X178$~& $abcadbeced$ &    0.7158~( 28) &    0.7230~(  9) &   -0.0072 &    0.7223~(  9) \\
$X179$~& $abcadbedce$ &   -0.4377~(  9) &   -0.4380~(  5) &    0.0003 &   -0.4379~(  4) \\
$X180$~& $abcadbedec$ &    0.0284~( 25) &    0.0265~(  9) &    0.0020 &    0.0267~(  9) \\
$X181$~& $abcadbeecd$ &   -4.4372~( 28) &   -4.4261~( 61) &   -0.0112 &   -4.4353~( 25) \\
$X182$~& $abcadbeedc$ &    1.2822~( 43) &    1.2771~( 50) &    0.0051 &    1.2800~( 33) \\
$X183$~& $abcadcbeed$ &   -0.0791~( 29) &   -0.0789~( 51) &   -0.0001 &   -0.0790~( 25) \\
$X184$~& $abcadcdeeb$ &    0.1973~(134) &    0.2284~(212) &   -0.0311 &    0.2062~(113) \\
$X185$~& $abcadcebed$ &   -0.1269~( 16) &   -0.1264~(  9) &   -0.0005 &   -0.1266~(  8) \\
$X186$~& $abcadcedeb$ &    1.1883~( 21) &    1.1905~(  9) &   -0.0022 &    1.1902~(  8) \\
$X187$~& $abcadceebd$ &    1.2699~( 27) &    1.2700~( 43) &   -0.0001 &    1.2699~( 23) \\
$X188$~& $abcadceedb$ &    1.7966~( 36) &    1.7937~( 72) &    0.0029 &    1.7960~( 32) \\
$X189$~& $abcaddbeec$ &   -3.7500~(105) &   -3.7574~(175) &    0.0073 &   -3.7520~( 90) \\
$X190$~& $abcaddceeb$ &   -2.4966~(217) &   -2.4741~(200) &   -0.0225 &   -2.4845~(147) \\
$X191$~& $abcaddebec$ &    0.1892~( 62) &    0.1892~( 69) &    0.0001 &    0.1892~( 46) \\
$X192$~& $abcaddeceb$ &    2.3868~( 91) &    2.3870~(180) &   -0.0003 &    2.3868~( 81) \\
$X193$~& $abcaddeebc$ &   -4.2570~( 84) &   -4.2686~(128) &    0.0116 &   -4.2605~( 70) \\
$X194$~& $abcaddeecb$ &   -0.6785~(102) &   -0.6797~(188) &    0.0012 &   -0.6787~( 89) \\
$X195$~& $abcadebcde$ &   -1.0708~( 20) &   -1.0706~(  9) &   -0.0002 &   -1.0706~(  8) \\
$X196$~& $abcadebced$ &   -2.0432~( 20) &   -2.0473~(  9) &    0.0040 &   -2.0466~(  8) \\
$X197$~& $abcadebdce$ &   -0.3848~(  8) &   -0.3838~(  4) &   -0.0010 &   -0.3840~(  3) \\
$X198$~& $abcadebdec$ &   -2.3533~( 26) &   -2.3583~(  9) &    0.0050 &   -2.3577~(  8) \\
$X199$~& $abcadebecd$ &    1.0636~( 26) &    1.0667~(  9) &   -0.0031 &    1.0664~(  8) \\
$X200$~& $abcadebedc$ &    0.0266~( 26) &    0.0259~(  9) &    0.0007 &    0.0260~(  9) \\
$X201$~& $abcadecbde$ &   -0.4897~( 18) &   -0.4887~(  8) &   -0.0010 &   -0.4889~(  7) \\
$X202$~& $abcadecbed$ &    1.9313~( 17) &    1.9363~(  9) &   -0.0050 &    1.9352~(  8) \\
$X203$~& $abcadecdbe$ &    0.9061~( 10) &    0.9075~(  7) &   -0.0014 &    0.9071~(  5) \\
$X204$~& $abcadecdeb$ &   -1.9485~( 26) &   -1.9449~(  9) &   -0.0036 &   -1.9453~(  8) \\
$X205$~& $abcadecebd$ &   -0.9039~( 13) &   -0.9044~(  8) &    0.0005 &   -0.9043~(  7) \\
$X206$~& $abcadecedb$ &    1.6836~( 23) &    1.6829~(  9) &    0.0007 &    1.6830~(  8) \\
$X207$~& $abcadedbce$ &    0.2908~( 23) &    0.2959~(  9) &   -0.0051 &    0.2952~(  8) \\
$X208$~& $abcadedbec$ &    0.5283~( 28) &    0.5312~(  9) &   -0.0028 &    0.5309~(  8) \\
$X209$~& $abcadedcbe$ &    0.1496~( 19) &    0.1511~(  8) &   -0.0015 &    0.1509~(  8) \\
$X210$~& $abcadedceb$ &    0.7803~( 19) &    0.7800~(  9) &    0.0003 &    0.7801~(  8) \\
$X211$~& $abcadedebc$ &    5.1339~( 90) &    5.1463~(129) &   -0.0124 &    5.1379~( 74) \\
$X212$~& $abcadedecb$ &   -0.4617~(138) &   -0.4539~(206) &   -0.0078 &   -0.4593~(114) \\
$X213$~& $abcadeebcd$ &   -2.4516~( 29) &   -2.4515~( 42) &   -0.0001 &   -2.4516~( 23) \\
$X214$~& $abcadeebdc$ &    0.6801~( 39) &    0.6777~( 69) &    0.0023 &    0.6795~( 34) \\
$X215$~& $abcadeecbd$ &    0.0724~( 24) &    0.0763~( 43) &   -0.0038 &    0.0734~( 21) \\
$X216$~& $abcadeecdb$ &   -1.3029~( 42) &   -1.3013~( 56) &   -0.0016 &   -1.3023~( 33) \\
$X217$~& $abcadeedbc$ &   -2.2261~( 71) &   -2.2348~(117) &    0.0086 &   -2.2285~( 60) \\
$X218$~& $abcadeedcb$ &   -1.6396~( 84) &   -1.6319~( 92) &   -0.0077 &   -1.6361~( 62) \\
$X219$~& $abcbaddeec$ &    1.3579~(311) &    1.3595~(178) &   -0.0015 &    1.3591~(155) \\
$X220$~& $abcbadedec$ &   -2.5734~(222) &   -2.5667~(214) &   -0.0068 &   -2.5699~(154) \\
$X221$~& $abcbadeecd$ &    0.6650~(161) &    0.6680~(206) &   -0.0030 &    0.6662~(127) \\
$X222$~& $abcbadeedc$ &    0.8293~(178) &    0.8025~(201) &    0.0267 &    0.8175~(133) \\
$X223$~& $abcbcdeeda$ &   17.5168~(349) &   17.5020~(156) &    0.0148 &   17.5045~(142) \\
$X224$~& $abcbdadeec$ &    2.4729~(110) &    2.5001~(196) &   -0.0273 &    2.4794~( 96) \\
$X225$~& $abcbdaedec$ &    0.3434~( 39) &    0.3419~( 12) &    0.0015 &    0.3421~( 12) \\
$X226$~& $abcbdaeecd$ &    1.0443~( 58) &    1.0390~( 77) &    0.0052 &    1.0424~( 46) \\
$X227$~& $abcbdaeedc$ &    0.5835~( 97) &    0.5991~(203) &   -0.0156 &    0.5864~( 87) \\
$X228$~& $abcbdceeda$ &   -6.8113~(333) &   -6.8120~(208) &    0.0007 &   -6.8118~(176) \\
$X229$~& $abcbddaeec$ &   -1.9843~(323) &   -1.9807~(210) &   -0.0036 &   -1.9818~(176) \\
$X230$~& $abcbddeeca$ &   15.6844~(350) &   15.6718~(199) &    0.0126 &   15.6749~(173) \\
$X231$~& $abcbdeadec$ &   -0.7737~( 28) &   -0.7723~( 10) &   -0.0014 &   -0.7725~( 10) \\
$X232$~& $abcbdeaedc$ &    0.4608~( 38) &    0.4604~( 12) &    0.0004 &    0.4605~( 11) \\
$X233$~& $abcbdecdea$ &    8.6698~(116) &    8.6613~(192) &    0.0085 &    8.6675~( 99) \\
$X234$~& $abcbdeceda$ &   -2.5793~(179) &   -2.5995~(210) &    0.0202 &   -2.5878~(136) \\
$X235$~& $abcbdedaec$ &    0.7486~( 35) &    0.7478~( 11) &    0.0009 &    0.7478~( 10) \\
$X236$~& $abcbdedcea$ &    2.0560~(180) &    2.1072~(205) &   -0.0512 &    2.0782~(135) \\
$X237$~& $abcbdedeca$ &  -12.9913~(363) &  -12.9686~(187) &   -0.0227 &  -12.9734~(166) \\
$X238$~& $abcbdeeadc$ &    1.2747~( 45) &    1.2837~( 92) &   -0.0090 &    1.2765~( 41) \\
$X239$~& $abcbdeecda$ &   -2.8075~(345) &   -2.8021~(201) &   -0.0053 &   -2.8035~(174) \\
$X240$~& $abcbdeedca$ &   10.9428~(298) &   10.9241~(209) &    0.0187 &   10.9303~(171) \\
$X241$~& $abccaddeeb$ &   13.8142~(357) &   13.7745~(201) &    0.0397 &   13.7841~(175) \\
$X242$~& $abccadedeb$ &  -10.4867~(377) &  -10.4478~(177) &   -0.0389 &  -10.4549~(160) \\
$X243$~& $abccadeedb$ &    3.8891~(336) &    3.8802~(199) &    0.0089 &    3.8825~(171) \\
$X244$~& $abccdadeeb$ &   -3.3041~(334) &   -3.2721~(187) &   -0.0320 &   -3.2797~(163) \\
$X245$~& $abccdaedeb$ &    0.0658~( 83) &    0.0880~(192) &   -0.0222 &    0.0693~( 76) \\
$X246$~& $abccdaeedb$ &   -0.3959~(174) &   -0.3816~(213) &   -0.0143 &   -0.3902~(134) \\
$X247$~& $abccddaeeb$ &   15.9539~(344) &   15.9573~(191) &   -0.0034 &   15.9565~(167) \\
$X248$~& $abccddeaeb$ &   -1.9165~(278) &   -1.9008~(209) &   -0.0157 &   -1.9065~(167) \\
$X249$~& $abccdeadeb$ &    4.0116~( 46) &    4.0143~( 66) &   -0.0027 &    4.0125~( 37) \\
$X250$~& $abccdeaedb$ &   -1.0558~( 68) &   -1.0478~(128) &   -0.0080 &   -1.0540~( 60) \\
$X251$~& $abccdedaeb$ &   -1.3906~( 76) &   -1.3435~(198) &   -0.0472 &   -1.3846~( 71) \\
$X252$~& $abccdedeab$ &  -10.9091~(343) &  -10.8565~(179) &   -0.0526 &  -10.8677~(158) \\
$X253$~& $abccdedeba$ &   17.8437~(352) &   17.8230~(196) &    0.0207 &   17.8279~(171) \\
$X254$~& $abccdeeadb$ &    2.2265~(175) &    2.2133~(217) &    0.0132 &    2.2213~(136) \\
$X255$~& $abccdeedab$ &    8.1598~(340) &    8.1520~(173) &    0.0078 &    8.1536~(154) \\
$X256$~& $abccdeedba$ &  -14.0405~(342) &  -13.9856~(194) &   -0.0549 &  -13.9990~(169) \\
$X257$~& $abcdabceed$ &    5.7475~( 51) &    5.7447~( 79) &    0.0029 &    5.7467~( 43) \\
$X258$~& $abcdabdeec$ &   -0.5254~( 39) &   -0.5246~( 55) &   -0.0008 &   -0.5252~( 32) \\
$X259$~& $abcdabeced$ &    0.0053~( 27) &    0.0050~( 10) &    0.0003 &    0.0050~(  9) \\
$X260$~& $abcdabedec$ &   -0.3958~( 20) &   -0.3927~(  8) &   -0.0031 &   -0.3932~(  8) \\
$X261$~& $abcdabeecd$ &    6.4046~( 30) &    6.3974~( 50) &    0.0072 &    6.4027~( 26) \\
$X262$~& $abcdabeedc$ &   -2.2854~( 24) &   -2.2848~( 38) &   -0.0005 &   -2.2852~( 20) \\
$X263$~& $abcdacbeed$ &   -2.8330~( 35) &   -2.8190~( 62) &   -0.0139 &   -2.8297~( 30) \\
$X264$~& $abcdacdeeb$ &    4.8826~( 64) &    4.8752~( 95) &    0.0074 &    4.8803~( 53) \\
$X265$~& $abcdacebed$ &   -0.6756~( 20) &   -0.6730~(  9) &   -0.0026 &   -0.6734~(  8) \\
$X266$~& $abcdacedeb$ &    0.1206~( 23) &    0.1225~(  9) &   -0.0019 &    0.1223~(  8) \\
$X267$~& $abcdaceebd$ &   -0.6608~( 19) &   -0.6591~( 31) &   -0.0017 &   -0.6603~( 16) \\
$X268$~& $abcdaceedb$ &    0.1185~( 31) &    0.1214~( 56) &   -0.0029 &    0.1192~( 27) \\
$X269$~& $abcdadbeec$ &   -0.7190~( 56) &   -0.7206~( 88) &    0.0016 &   -0.7195~( 47) \\
$X270$~& $abcdadceeb$ &   -1.6881~( 97) &   -1.6705~(217) &   -0.0176 &   -1.6851~( 88) \\
$X271$~& $abcdadebec$ &    0.2492~( 23) &    0.2505~(  9) &   -0.0012 &    0.2503~(  8) \\
$X272$~& $abcdadeceb$ &   -0.7285~( 32) &   -0.7297~( 11) &    0.0012 &   -0.7296~( 10) \\
$X273$~& $abcdadeebc$ &   -2.0474~( 45) &   -2.0422~( 75) &   -0.0052 &   -2.0460~( 39) \\
$X274$~& $abcdadeecb$ &    0.8675~( 72) &    0.8768~(117) &   -0.0093 &    0.8701~( 61) \\
$X275$~& $abcdaebced$ &   -0.7496~( 12) &   -0.7478~(  8) &   -0.0018 &   -0.7484~(  7) \\
$X276$~& $abcdaebdce$ &   -0.5547~( 10) &   -0.5540~(  5) &   -0.0007 &   -0.5541~(  4) \\
$X277$~& $abcdaebdec$ &    2.7936~( 10) &    2.7944~(  6) &   -0.0008 &    2.7942~(  5) \\
$X278$~& $abcdaebecd$ &   -0.1577~( 23) &   -0.1602~(  9) &    0.0025 &   -0.1598~(  9) \\
$X279$~& $abcdaebedc$ &    0.8399~( 15) &    0.8408~(  8) &   -0.0009 &    0.8406~(  7) \\
$X280$~& $abcdaecbed$ &   -1.0127~(  8) &   -1.0114~(  5) &   -0.0013 &   -1.0118~(  4) \\
$X281$~& $abcdaecdeb$ &   -1.3732~( 25) &   -1.3703~(  9) &   -0.0030 &   -1.3707~(  9) \\
$X282$~& $abcdaecebd$ &    0.4907~( 18) &    0.4922~(  9) &   -0.0015 &    0.4919~(  8) \\
$X283$~& $abcdaecedb$ &   -0.0427~( 23) &   -0.0396~(  8) &   -0.0030 &   -0.0400~(  8) \\
$X284$~& $abcdaedbec$ &   -0.2670~(  9) &   -0.2673~(  5) &    0.0003 &   -0.2673~(  4) \\
$X285$~& $abcdaedceb$ &    0.0271~( 16) &    0.0277~(  7) &   -0.0007 &    0.0276~(  6) \\
$X286$~& $abcdaedebc$ &    0.8014~( 21) &    0.8041~(  9) &   -0.0026 &    0.8036~(  8) \\
$X287$~& $abcdaedecb$ &    0.2014~( 19) &    0.2021~(  9) &   -0.0008 &    0.2020~(  8) \\
$X288$~& $abcdaeebcd$ &    4.2112~( 28) &    4.2137~( 45) &   -0.0025 &    4.2119~( 24) \\
$X289$~& $abcdaeebdc$ &   -1.5651~( 19) &   -1.5662~( 35) &    0.0011 &   -1.5654~( 17) \\
$X290$~& $abcdaeecbd$ &   -3.7763~( 23) &   -3.7736~( 42) &   -0.0027 &   -3.7756~( 20) \\
$X291$~& $abcdaeecdb$ &    1.5957~( 32) &    1.5890~( 62) &    0.0067 &    1.5943~( 28) \\
$X292$~& $abcdaeedbc$ &    0.9114~( 36) &    0.9144~( 49) &   -0.0030 &    0.9125~( 29) \\
$X293$~& $abcdaeedcb$ &   -1.2653~( 41) &   -1.2582~( 58) &   -0.0070 &   -1.2629~( 33) \\
$X294$~& $abcdbaceed$ &   -3.3891~( 25) &   -3.3873~( 36) &   -0.0018 &   -3.3885~( 20) \\
$X295$~& $abcdbadeec$ &    1.7883~( 26) &    1.7913~( 44) &   -0.0030 &    1.7891~( 22) \\
$X296$~& $abcdbaeced$ &    0.5511~( 13) &    0.5528~(  8) &   -0.0017 &    0.5522~(  7) \\
$X297$~& $abcdbaedec$ &   -0.4696~( 16) &   -0.4693~(  9) &   -0.0003 &   -0.4694~(  7) \\
$X298$~& $abcdbaeecd$ &   -1.9142~( 28) &   -1.9153~( 44) &    0.0011 &   -1.9145~( 23) \\
$X299$~& $abcdbaeedc$ &   -0.2907~( 22) &   -0.2887~( 39) &   -0.0020 &   -0.2902~( 19) \\
$X300$~& $abcdbceeda$ &   -9.4327~(194) &   -9.4309~(210) &   -0.0018 &   -9.4318~(142) \\
$X301$~& $abcdbdaeec$ &   -1.3351~( 81) &   -1.3445~(122) &    0.0094 &   -1.3380~( 68) \\
$X302$~& $abcdbdeeca$ &   -1.8294~(223) &   -1.8502~(216) &    0.0208 &   -1.8401~(155) \\
$X303$~& $abcdbeadec$ &    0.3341~(  7) &    0.3348~(  3) &   -0.0007 &    0.3347~(  3) \\
$X304$~& $abcdbeaecd$ &   -0.3397~( 16) &   -0.3381~(  9) &   -0.0016 &   -0.3385~(  8) \\
$X305$~& $abcdbeaedc$ &    0.4715~( 14) &    0.4719~(  7) &   -0.0004 &    0.4718~(  6) \\
$X306$~& $abcdbeceda$ &    0.1228~( 55) &    0.1167~( 86) &    0.0062 &    0.1210~( 46) \\
$X307$~& $abcdbedeca$ &   -0.3071~( 59) &   -0.3024~(107) &   -0.0048 &   -0.3060~( 52) \\
$X308$~& $abcdbeeadc$ &    1.8122~( 22) &    1.8126~( 40) &   -0.0004 &    1.8123~( 19) \\
$X309$~& $abcdbeecda$ &   -4.2448~(173) &   -4.2500~(213) &    0.0051 &   -4.2469~(134) \\
$X310$~& $abcdbeedca$ &    0.2490~(191) &    0.2397~(211) &    0.0093 &    0.2448~(142) \\
$X311$~& $abcdcabeed$ &   -0.5291~( 58) &   -0.5389~( 78) &    0.0098 &   -0.5326~( 47) \\
$X312$~& $abcdcadeeb$ &   -1.2454~(139) &   -1.2693~( 90) &    0.0239 &   -1.2622~( 76) \\
$X313$~& $abcdcaebed$ &    0.9660~( 38) &    0.9654~( 12) &    0.0006 &    0.9654~( 11) \\
$X314$~& $abcdcaedeb$ &    0.8266~( 29) &    0.8335~( 11) &   -0.0069 &    0.8327~( 10) \\
$X315$~& $abcdcaeebd$ &   -1.3728~( 43) &   -1.3787~( 67) &    0.0059 &   -1.3745~( 36) \\
$X316$~& $abcdcaeedb$ &    0.0094~( 39) &    0.0272~( 89) &   -0.0178 &    0.0123~( 36) \\
$X317$~& $abcdcbeeda$ &    1.4535~(221) &    1.4828~(204) &   -0.0293 &    1.4693~(150) \\
$X318$~& $abcdcdaeeb$ &   -8.7568~(343) &   -8.7479~(201) &   -0.0089 &   -8.7502~(174) \\
$X319$~& $abcdcdeaeb$ &    0.6801~(179) &    0.6449~(213) &    0.0352 &    0.6655~(137) \\
$X320$~& $abcdceadeb$ &    0.5627~( 17) &    0.5641~(  9) &   -0.0014 &    0.5637~(  8) \\
$X321$~& $abcdceaedb$ &   -0.9005~( 26) &   -0.8961~( 10) &   -0.0044 &   -0.8967~(  9) \\
$X322$~& $abcdcedaeb$ &    0.9338~( 23) &    0.9364~(  9) &   -0.0025 &    0.9360~(  9) \\
$X323$~& $abcdceeadb$ &   -0.0053~( 40) &    0.0092~( 85) &   -0.0145 &   -0.0026~( 36) \\
$X324$~& $abcdceedab$ &   -8.8058~(243) &   -8.8139~(209) &    0.0081 &   -8.8105~(158) \\
$X325$~& $abcdceedba$ &   11.5958~(343) &   11.5456~(198) &    0.0503 &   11.5582~(172) \\
$X326$~& $abcddabeec$ &   -9.0047~(145) &   -8.9830~(204) &   -0.0217 &   -8.9974~(118) \\
$X327$~& $abcddaceeb$ &    1.5517~(229) &    1.5868~(209) &   -0.0351 &    1.5709~(154) \\
$X328$~& $abcddaebec$ &   -0.2781~( 42) &   -0.2745~( 74) &   -0.0035 &   -0.2772~( 36) \\
$X329$~& $abcddaeceb$ &   -0.9627~( 67) &   -0.9506~(103) &   -0.0121 &   -0.9591~( 56) \\
$X330$~& $abcddaeebc$ &   -4.9591~( 88) &   -4.9451~(125) &   -0.0141 &   -4.9545~( 72) \\
$X331$~& $abcddaeecb$ &    4.7241~(127) &    4.7164~(215) &    0.0077 &    4.7221~(109) \\
$X332$~& $abcddbaeec$ &    3.0539~(161) &    3.0436~(206) &    0.0103 &    3.0500~(127) \\
$X333$~& $abcddbeeca$ &    6.8088~(341) &    6.8339~(194) &   -0.0251 &    6.8278~(169) \\
$X334$~& $abcddcaeeb$ &    5.1727~(340) &    5.1696~(189) &    0.0031 &    5.1703~(165) \\
$X335$~& $abcddceaeb$ &   -2.0294~(132) &   -2.0421~(212) &    0.0127 &   -2.0329~(112) \\
$X336$~& $abcddeabec$ &   -0.7685~( 20) &   -0.7730~( 30) &    0.0045 &   -0.7700~( 17) \\
$X337$~& $abcddeaceb$ &   -1.2039~( 32) &   -1.1991~( 55) &   -0.0048 &   -1.2027~( 27) \\
$X338$~& $abcddeaebc$ &   -1.8505~( 38) &   -1.8492~( 69) &   -0.0012 &   -1.8502~( 33) \\
$X339$~& $abcddeaecb$ &    0.4111~( 40) &    0.4151~( 59) &   -0.0040 &    0.4124~( 33) \\
$X340$~& $abcddebeca$ &   -2.1543~(202) &   -2.1472~(214) &   -0.0071 &   -2.1509~(147) \\
$X341$~& $abcddecaeb$ &    1.7815~( 33) &    1.7782~( 65) &    0.0033 &    1.7808~( 30) \\
$X342$~& $abcddeeacb$ &    2.6063~(125) &    2.6128~(115) &   -0.0065 &    2.6099~( 84) \\
$X343$~& $abcdeabced$ &    3.8873~( 30) &    3.8962~( 10) &   -0.0089 &    3.8952~(  9) \\
$X344$~& $abcdeabdce$ &    3.4223~( 18) &    3.4239~(  9) &   -0.0016 &    3.4236~(  8) \\
$X345$~& $abcdeabdec$ &   -1.0075~( 18) &   -1.0069~(  9) &   -0.0006 &   -1.0070~(  8) \\
$X346$~& $abcdeabecd$ &    0.2864~( 20) &    0.2904~(  9) &   -0.0041 &    0.2898~(  8) \\
$X347$~& $abcdeabedc$ &   -2.6846~( 21) &   -2.6875~(  9) &    0.0029 &   -2.6870~(  9) \\
$X348$~& $abcdeacbed$ &   -0.4899~( 15) &   -0.4905~(  9) &    0.0005 &   -0.4903~(  7) \\
$X349$~& $abcdeacdeb$ &    2.0800~( 36) &    2.0793~(  9) &    0.0007 &    2.0794~(  9) \\
$X350$~& $abcdeacebd$ &    1.4643~( 11) &    1.4649~(  6) &   -0.0007 &    1.4648~(  5) \\
$X351$~& $abcdeacedb$ &    0.2554~( 20) &    0.2536~(  8) &    0.0017 &    0.2539~(  8) \\
$X352$~& $abcdeadbec$ &   -0.1260~(  8) &   -0.1257~(  4) &   -0.0003 &   -0.1258~(  4) \\
$X353$~& $abcdeadceb$ &    0.1950~( 16) &    0.1952~(  8) &   -0.0002 &    0.1952~(  7) \\
$X354$~& $abcdeadebc$ &   -2.0503~( 20) &   -2.0501~(  9) &   -0.0002 &   -2.0501~(  8) \\
$X355$~& $abcdeadecb$ &   -1.0738~( 25) &   -1.0747~(  9) &    0.0009 &   -1.0746~(  8) \\
$X356$~& $abcdeaebcd$ &    2.0684~( 24) &    2.0685~(  9) &   -0.0002 &    2.0685~(  9) \\
$X357$~& $abcdeaebdc$ &    0.3746~( 16) &    0.3760~(  8) &   -0.0014 &    0.3757~(  7) \\
$X358$~& $abcdeaecbd$ &    0.0463~( 16) &    0.0474~(  8) &   -0.0011 &    0.0472~(  7) \\
$X359$~& $abcdeaecdb$ &   -0.1396~( 17) &   -0.1381~(  9) &   -0.0015 &   -0.1384~(  8) \\
$X360$~& $abcdeaedbc$ &   -0.4604~( 37) &   -0.4592~( 10) &   -0.0012 &   -0.4593~( 10) \\
$X361$~& $abcdeaedcb$ &    2.5600~( 26) &    2.5629~(  9) &   -0.0029 &    2.5625~(  8) \\
$X362$~& $abcdebadec$ &   -0.5714~( 12) &   -0.5729~(  8) &    0.0014 &   -0.5724~(  6) \\
$X363$~& $abcdebaecd$ &   -2.3442~( 19) &   -2.3475~(  9) &    0.0033 &   -2.3468~(  8) \\
$X364$~& $abcdebaedc$ &    2.3957~( 18) &    2.4006~(  9) &   -0.0049 &    2.3995~(  8) \\
$X365$~& $abcdebceda$ &    0.4177~( 30) &    0.4187~( 44) &   -0.0011 &    0.4180~( 24) \\
$X366$~& $abcdebdeca$ &    5.6759~( 43) &    5.6790~(110) &   -0.0031 &    5.6763~( 40) \\
$X367$~& $abcdebeadc$ &   -0.7176~( 12) &   -0.7168~(  8) &   -0.0008 &   -0.7170~(  7) \\
$X368$~& $abcdebecda$ &   -0.3404~( 45) &   -0.3420~( 79) &    0.0015 &   -0.3408~( 39) \\
$X369$~& $abcdebedca$ &   -3.3812~( 59) &   -3.3665~(121) &   -0.0147 &   -3.3783~( 53) \\
$X370$~& $abcdecadeb$ &   -1.4763~( 12) &   -1.4741~(  7) &   -0.0022 &   -1.4747~(  6) \\
$X371$~& $abcdecaedb$ &    0.0045~( 10) &    0.0050~(  4) &   -0.0005 &    0.0049~(  4) \\
$X372$~& $abcdecdaeb$ &   -1.2900~( 33) &   -1.2913~(  9) &    0.0013 &   -1.2912~(  9) \\
$X373$~& $abcdeceadb$ &    0.5851~( 24) &    0.5877~(  9) &   -0.0025 &    0.5874~(  8) \\
$X374$~& $abcdecedab$ &    0.9188~(266) &    0.9318~(166) &   -0.0130 &    0.9281~(141) \\
$X375$~& $abcdecedba$ &    1.0991~(163) &    1.0880~(210) &    0.0111 &    1.0949~(129) \\
$X376$~& $abcdedabec$ &    1.0484~( 16) &    1.0514~(  7) &   -0.0030 &    1.0509~(  6) \\
$X377$~& $abcdedaceb$ &    0.4264~( 27) &    0.4313~(  9) &   -0.0049 &    0.4307~(  8) \\
$X378$~& $abcdedaebc$ &    1.3196~( 21) &    1.3238~(  9) &   -0.0042 &    1.3232~(  8) \\
$X379$~& $abcdedaecb$ &   -0.3201~( 17) &   -0.3198~(  9) &   -0.0003 &   -0.3198~(  8) \\
$X380$~& $abcdedbeca$ &   -1.0268~( 48) &   -1.0216~( 91) &   -0.0052 &   -1.0257~( 43) \\
$X381$~& $abcdedcaeb$ &    1.0861~( 29) &    1.0882~(  9) &   -0.0022 &    1.0880~(  9) \\
$X382$~& $abcdedeacb$ &   -1.7712~( 80) &   -1.7582~(142) &   -0.0130 &   -1.7681~( 70) \\
$X383$~& $abcdeeabdc$ &   -4.8034~( 22) &   -4.7978~( 35) &   -0.0056 &   -4.8018~( 19) \\
$X384$~& $abcdeeacdb$ &    1.9266~( 31) &    1.9384~( 57) &   -0.0118 &    1.9293~( 27) \\
$X385$~& $abcdeeadbc$ &   -0.7427~( 19) &   -0.7408~( 30) &   -0.0019 &   -0.7421~( 16) \\
$X386$~& $abcdeeadcb$ &    0.6887~( 38) &    0.6877~( 59) &    0.0010 &    0.6884~( 32) \\
$X387$~& $abcdeebdca$ &    1.9508~(152) &    1.9763~(208) &   -0.0255 &    1.9597~(123) \\
$X388$~& $abcdeecadb$ &   -0.4349~( 40) &   -0.4336~( 49) &   -0.0013 &   -0.4344~( 31) \\
$X389$~& $abcdeedacb$ &   -0.0433~( 68) &   -0.0525~(123) &    0.0092 &   -0.0455~( 59) \\
\end{longtable*}
\endgroup

%% file: table_residual.tex
\renewcommand{\baselinestretch}{0.95}
\begin{table*}
\caption{
Residual renormalization constants used to calculate
$A_1^{(10)} [\text{Set~V}]$. 
The  $\Delta M_n$, $\Delta L\!B_n$, and $\Delta dm_n$ are the sum of the finite magnetic moment
amplitudes, the sum of the finite parts of vertex and
wave-function renormalization constants, and the sum of the finite parts of the mass-renormalization constants,
respectively, 
all derived from the $n$th order-diagrams without a fermion loop 
of the QED perturbation theory.
$\Delta M_{10}$ is newly calculated in this work. $\Delta M_{8}$ is derived from the near-exact result
Eq.~(5) of Ref.\cite{Laporta:2017okg}. Other finite integrals are copied from the previous work TABLE~II of Ref.~\cite{Aoyama:2014sxa}.
 \label{table:residual}
}
\hfill
\begin{minipage}[t][][b]{.49\textwidth}
\begin{ruledtabular}
\begin{tabular}{@{\hskip1em}l@{}d@{\hskip1em}}
\multicolumn{1}{c}{Integral} &
\multicolumn{1}{c}{Value (Error)} \\
\hline
    $ \Delta M_{10}$  &        2.350~(192)          \\
    $ \Delta M_{8}$   &        1.738~467~(20)        \\
    $ \Delta M_{6}$   &        0.425~8135~(30)      \\
    $ \Delta M_{4}$   &        0.030~833~612\cdots  \\
    $  M_{2}$         &        0.5                  \\
    $ \Delta L\!B_{8}$  &        2.0504~(86)        \\ 
    $ \Delta L\!B_{6}$  &        0.100~801~(43)     \\  
    $ \Delta L\!B_{4}$  &        0.027~9171~(61)    \\ 
    $ \Delta L\!B_{2}$  &        0.75               \\
\end{tabular}
\end{ruledtabular}
\end{minipage}
%
\hfill
\begin{minipage}[t][][b]{.49\textwidth}
\begin{ruledtabular}
\begin{tabular}{@{\hskip1em}l@{}d@{\hskip1em}}
\multicolumn{1}{c}{Integral} &
\multicolumn{1}{c}{Value (Error)} \\
\hline
    $ \Delta L_{4^*}$ &       -0.459~051~(62)         \\ 
    $ \Delta L_{2^*}$ &       -0.75                   \\
    $ \Delta dm_{6}$  &       -2.340~815~(55)         \\ 
    $ \Delta dm_{4}$  &        1.906~3609~(90)        \\
    $ \Delta dm_{2^*}$&       -0.75                   \\
\\
\\
\\
\\
\end{tabular}
\end{ruledtabular}
\end{minipage}
\hfill
\end{table*}

\renewcommand{\baselinestretch}{1.0}

%% file: table_DLB8.tex
\renewcommand{\baselinestretch}{0.95}
\begin{table*}
\caption{
Integrals contributing to the residual renormalization
constants $\DLB_8$.
The 47 10-dimensional integrals derived from the eighth-order
vertex and self-energy renoromalization constants 
are evaluated by VEGAS.
The second column shows the diagarm representation by
photon labels attached to the fermion line of a self-energy diagram.
For $\LB04, \LB11, \LB12, \LB16, \LB17, \LB18, \LB29$, and $\LB30$,
double-double precision for real numbers is partially used. 
Others are evaluated in double precision.
 \label{table:DLB8}
}
\hfill
\begin{minipage}[t][][b]{.49\textwidth}
\begin{ruledtabular}
\begin{tabular}{@{\hskip 1em}l@{\hskip 1em}l@{\hskip -3em }d@{\hskip 2em}}
\multicolumn{1}{l}{Integral} &
\multicolumn{1}{@{\hskip 0em}l}{Expression} &
\multicolumn{1}{@{\hskip 0em}c}{Value (Error)} \\
\hline
$\LB01$ & $abacbdcd$ &      -0.190~96~( 80) \\
$\LB02$ & $abacbddc$ &       0.604~07~(177) \\
$\LB03$ & $abaccdbd$ &       0.427~96~(150) \\
$\LB04$ & $abaccddb$ &      -1.005~69~(198) \\
$\LB05$ & $abacdbcd$ &      -0.819~62~( 71) \\
$\LB06$ & $abacdbdc$ &      -0.101~17~( 96) \\
$\LB07$ & $abacdcbd$ &      -0.165~26~( 94) \\
$\LB08$ & $abacdcdb$ &       0.473~20~(185) \\
$\LB09$ & $abacddbc$ &       0.773~21~(187) \\
$\LB10$ & $abacddcb$ &      -0.484~03~(177) \\
$\LB11$ & $abbcaddc$ &      -0.403~21~(185) \\
$\LB12$ & $abbccdda$ &       0.606~68~(189) \\
$\LB13$ & $abbcdacd$ &       0.904~58~(109) \\
$\LB14$ & $abbcdadc$ &       0.164~56~(142) \\
$\LB15$ & $abbcdcad$ &       0.136~19~(136) \\
$\LB16$ & $abbcdcda$ &      -1.013~14~(239) \\
$\LB17$ & $abbcddac$ &      -0.975~11~(193) \\
$\LB18$ & $abbcddca$ &       3.204~32~(195) \\
$\LB19$ & $abcadbcd$ &       0.217~77~( 18) \\
$\LB20$ & $abcadbdc$ &      -0.445~27~( 40) \\
$\LB21$ & $abcadcbd$ &      -0.168~93~( 11) \\
$\LB22$ & $abcadcdb$ &       0.177~41~( 40) \\
$\LB23$ & $abcaddbc$ &       0.602~13~(115) \\
$\LB24$ & $abcaddcb$ &       0.067~22~( 87) \\
\end{tabular}
\end{ruledtabular}
\end{minipage}
%
\hfill
\begin{minipage}[t][][b]{.49\textwidth}
\begin{ruledtabular}
\begin{tabular}{@{\hskip 1em}l@{\hskip 1em}l@{\hskip -3em }d@{\hskip 2em}}
\multicolumn{1}{l}{Integral} &
\multicolumn{1}{@{\hskip 0em}l}{Expression} &
\multicolumn{1}{@{\hskip 0em}c}{Value (Error)} \\
\hline
$\LB25$ & $abcbdadc$ &       0.013~43~( 23) \\
$\LB26$ & $abcbdcda$ &       0.351~36~( 94) \\
$\LB27$ & $abcbddac$ &       0.260~02~(130) \\
$\LB28$ & $abcbddca$ &      -0.858~51~(231) \\
$\LB29$ & $abccddab$ &      -0.491~48~(145) \\
$\LB30$ & $abccddba$ &       0.496~81~(217) \\
$\LB31$ & $abcdabcd$ &      -0.831~79~( 28) \\
$\LB32$ & $abcdabdc$ &       0.387~67~( 29) \\
$\LB33$ & $abcdacbd$ &       0.259~53~( 11) \\
$\LB34$ & $abcdacdb$ &      -0.312~91~( 30) \\
$\LB35$ & $abcdadbc$ &      -0.369~11~( 36) \\
$\LB36$ & $abcdadcb$ &       0.077~06~( 48) \\
$\LB37$ & $abcdbadc$ &      -0.174~46~( 21) \\
$\LB38$ & $abcdbcda$ &       0.086~96~( 37) \\
$\LB39$ & $abcdbdac$ &       0.220~33~( 33) \\
$\LB40$ & $abcdbdca$ &       0.179~34~(183) \\
$\LB41$ & $abcdcdab$ &       0.292~11~( 58) \\
$\LB42$ & $abcdcdba$ &      -0.310~87~( 91) \\
$\LB43$ & $abcddabc$ &       0.629~11~( 39) \\
$\LB44$ & $abcddacb$ &      -0.154~60~(108) \\
$\LB45$ & $abcddbca$ &      -0.296~17~(106) \\
$\LB46$ & $abcddcab$ &      -0.325~17~( 66) \\
$\LB47$ & $abcddcba$ &       0.334~83~(116) \\
\end{tabular}
\end{ruledtabular}
\end{minipage}
\hfill
\end{table*}

\renewcommand{\baselinestretch}{1.0}

%% file: setV2017.bbl
\begin{thebibliography}{97}
\expandafter\ifx\csname natexlab\endcsname\relax\def\natexlab#1{#1}\fi
\expandafter\ifx\csname bibnamefont\endcsname\relax
  \def\bibnamefont#1{#1}\fi
\expandafter\ifx\csname bibfnamefont\endcsname\relax
  \def\bibfnamefont#1{#1}\fi
\expandafter\ifx\csname citenamefont\endcsname\relax
  \def\citenamefont#1{#1}\fi
\expandafter\ifx\csname url\endcsname\relax
  \def\url#1{\texttt{#1}}\fi
\expandafter\ifx\csname urlprefix\endcsname\relax\def\urlprefix{URL }\fi
\providecommand{\bibinfo}[2]{#2}
\providecommand{\eprint}[2][]{\url{#2}}

\bibitem[{\citenamefont{Kusch and Foley}(1947)}]{Kusch:1947}
\bibinfo{author}{\bibfnamefont{P.}~\bibnamefont{Kusch}} \bibnamefont{and}
  \bibinfo{author}{\bibfnamefont{H.~M.} \bibnamefont{Foley}},
  \bibinfo{journal}{Phys. Rev.} \textbf{\bibinfo{volume}{72}},
  \bibinfo{pages}{1256} (\bibinfo{year}{1947}).

\bibitem[{\citenamefont{Schwinger}(1948)}]{Schwinger:1948iu}
\bibinfo{author}{\bibfnamefont{J.~S.} \bibnamefont{Schwinger}},
  \bibinfo{journal}{Phys. Rev.} \textbf{\bibinfo{volume}{73}},
  \bibinfo{pages}{416} (\bibinfo{year}{1948}).

\bibitem[{\citenamefont{Hanneke et~al.}(2008)\citenamefont{Hanneke, Fogwell,
  and Gabrielse}}]{Hanneke:2008tm}
\bibinfo{author}{\bibfnamefont{D.}~\bibnamefont{Hanneke}},
  \bibinfo{author}{\bibfnamefont{S.}~\bibnamefont{Fogwell}}, \bibnamefont{and}
  \bibinfo{author}{\bibfnamefont{G.}~\bibnamefont{Gabrielse}},
  \bibinfo{journal}{Phys. Rev. Lett.} \textbf{\bibinfo{volume}{100}},
  \bibinfo{pages}{120801} (\bibinfo{year}{2008}).

\bibitem[{\citenamefont{Hanneke et~al.}(2011)\citenamefont{Hanneke,
  Fogwell~Hoogerheide, and Gabrielse}}]{Hanneke:2010au}
\bibinfo{author}{\bibfnamefont{D.}~\bibnamefont{Hanneke}},
  \bibinfo{author}{\bibfnamefont{S.}~\bibnamefont{Fogwell~Hoogerheide}},
  \bibnamefont{and}
  \bibinfo{author}{\bibfnamefont{G.}~\bibnamefont{Gabrielse}},
  \bibinfo{journal}{Phys. Rev. A} \textbf{\bibinfo{volume}{83}},
  \bibinfo{pages}{052122} (\bibinfo{year}{2011}).

\bibitem[{\citenamefont{Van~Dyck et~al.}(1987)\citenamefont{Van~Dyck,
  Schwinberg, and Dehmelt}}]{VanDyck:1987ay}
\bibinfo{author}{\bibfnamefont{R.~S.} \bibnamefont{Van~Dyck}},
  \bibinfo{author}{\bibfnamefont{P.~B.} \bibnamefont{Schwinberg}},
  \bibnamefont{and} \bibinfo{author}{\bibfnamefont{H.~G.}
  \bibnamefont{Dehmelt}}, \bibinfo{journal}{Phys. Rev. Lett.}
  \textbf{\bibinfo{volume}{59}}, \bibinfo{pages}{26} (\bibinfo{year}{1987}).

\bibitem[{\citenamefont{Fogwell~Hoogerheide
  et~al.}(2015)\citenamefont{Fogwell~Hoogerheide, Dorr, Novitski, and
  Gabrielse}}]{Hoogerheide:2014mna}
\bibinfo{author}{\bibfnamefont{S.}~\bibnamefont{Fogwell~Hoogerheide}},
  \bibinfo{author}{\bibfnamefont{J.~C.} \bibnamefont{Dorr}},
  \bibinfo{author}{\bibfnamefont{E.}~\bibnamefont{Novitski}}, \bibnamefont{and}
  \bibinfo{author}{\bibfnamefont{G.}~\bibnamefont{Gabrielse}},
  \bibinfo{journal}{Rev. Sci. Instrum.} \textbf{\bibinfo{volume}{86}},
  \bibinfo{pages}{053301} (\bibinfo{year}{2015}).

\bibitem[{\citenamefont{Elend}(1966)}]{Elend:1966a}
\bibinfo{author}{\bibfnamefont{H.~H.} \bibnamefont{Elend}},
  \bibinfo{journal}{Phys. Lett.} \textbf{\bibinfo{volume}{20}},
  \bibinfo{pages}{682} (\bibinfo{year}{1966}).

\bibitem[{\citenamefont{Samuel and Li}(1991)}]{Samuel:1990qf}
\bibinfo{author}{\bibfnamefont{M.~A.} \bibnamefont{Samuel}} \bibnamefont{and}
  \bibinfo{author}{\bibfnamefont{G.-w.} \bibnamefont{Li}},
  \bibinfo{journal}{Phys. Rev. D} \textbf{\bibinfo{volume}{44}},
  \bibinfo{pages}{3935} (\bibinfo{year}{1991}).

\bibitem[{\citenamefont{Li et~al.}(1993)\citenamefont{Li, Mendel, and
  Samuel}}]{Li:1992xf}
\bibinfo{author}{\bibfnamefont{G.}~\bibnamefont{Li}},
  \bibinfo{author}{\bibfnamefont{R.}~\bibnamefont{Mendel}}, \bibnamefont{and}
  \bibinfo{author}{\bibfnamefont{M.~A.} \bibnamefont{Samuel}},
  \bibinfo{journal}{Phys. Rev. D} \textbf{\bibinfo{volume}{47}},
  \bibinfo{pages}{1723} (\bibinfo{year}{1993}).

\bibitem[{\citenamefont{Laporta and Remiddi}(1993)}]{Laporta:1992pa}
\bibinfo{author}{\bibfnamefont{S.}~\bibnamefont{Laporta}} \bibnamefont{and}
  \bibinfo{author}{\bibfnamefont{E.}~\bibnamefont{Remiddi}},
  \bibinfo{journal}{Phys. Lett.} \textbf{\bibinfo{volume}{B301}},
  \bibinfo{pages}{440} (\bibinfo{year}{1993}).

\bibitem[{\citenamefont{Laporta}(1993)}]{Laporta:1993ju}
\bibinfo{author}{\bibfnamefont{S.}~\bibnamefont{Laporta}},
  \bibinfo{journal}{Nuovo Cim.} \textbf{\bibinfo{volume}{A106}},
  \bibinfo{pages}{675} (\bibinfo{year}{1993}).

\bibitem[{\citenamefont{Passera}(2007)}]{Passera:2006gc}
\bibinfo{author}{\bibfnamefont{M.}~\bibnamefont{Passera}},
  \bibinfo{journal}{Phys. Rev. D} \textbf{\bibinfo{volume}{75}},
  \bibinfo{pages}{013002} (\bibinfo{year}{2007}).

\bibitem[{\citenamefont{Kinoshita and Nio}(2006)}]{Kinoshita:2005sm}
\bibinfo{author}{\bibfnamefont{T.}~\bibnamefont{Kinoshita}} \bibnamefont{and}
  \bibinfo{author}{\bibfnamefont{M.}~\bibnamefont{Nio}},
  \bibinfo{journal}{Phys. Rev. D} \textbf{\bibinfo{volume}{73}},
  \bibinfo{pages}{053007} (\bibinfo{year}{2006}).

\bibitem[{\citenamefont{Kataev}(2012)}]{Kataev:2012kn}
\bibinfo{author}{\bibfnamefont{A.~L.} \bibnamefont{Kataev}},
  \bibinfo{journal}{Phys.Rev.} \textbf{\bibinfo{volume}{D86}},
  \bibinfo{pages}{013010} (\bibinfo{year}{2012}).

\bibitem[{\citenamefont{Kurz et~al.}(2014{\natexlab{a}})\citenamefont{Kurz,
  Liu, Marquard, and Steinhauser}}]{Kurz:2013exa}
\bibinfo{author}{\bibfnamefont{A.}~\bibnamefont{Kurz}},
  \bibinfo{author}{\bibfnamefont{T.}~\bibnamefont{Liu}},
  \bibinfo{author}{\bibfnamefont{P.}~\bibnamefont{Marquard}}, \bibnamefont{and}
  \bibinfo{author}{\bibfnamefont{M.}~\bibnamefont{Steinhauser}},
  \bibinfo{journal}{Nucl. Phys.} \textbf{\bibinfo{volume}{B879}},
  \bibinfo{pages}{1} (\bibinfo{year}{2014}{\natexlab{a}}).

\bibitem[{\citenamefont{Aoyama et~al.}(2008{\natexlab{a}})\citenamefont{Aoyama,
  Hayakawa, Kinoshita, Nio, and Watanabe}}]{Aoyama:2008gy}
\bibinfo{author}{\bibfnamefont{T.}~\bibnamefont{Aoyama}},
  \bibinfo{author}{\bibfnamefont{M.}~\bibnamefont{Hayakawa}},
  \bibinfo{author}{\bibfnamefont{T.}~\bibnamefont{Kinoshita}},
  \bibinfo{author}{\bibfnamefont{M.}~\bibnamefont{Nio}}, \bibnamefont{and}
  \bibinfo{author}{\bibfnamefont{N.}~\bibnamefont{Watanabe}},
  \bibinfo{journal}{Phys. Rev. D} \textbf{\bibinfo{volume}{78}},
  \bibinfo{pages}{053005} (\bibinfo{year}{2008}{\natexlab{a}}).

\bibitem[{\citenamefont{Aoyama et~al.}(2008{\natexlab{b}})\citenamefont{Aoyama,
  Hayakawa, Kinoshita, and Nio}}]{Aoyama:2008hz}
\bibinfo{author}{\bibfnamefont{T.}~\bibnamefont{Aoyama}},
  \bibinfo{author}{\bibfnamefont{M.}~\bibnamefont{Hayakawa}},
  \bibinfo{author}{\bibfnamefont{T.}~\bibnamefont{Kinoshita}},
  \bibnamefont{and} \bibinfo{author}{\bibfnamefont{M.}~\bibnamefont{Nio}},
  \bibinfo{journal}{Phys. Rev. D} \textbf{\bibinfo{volume}{78}},
  \bibinfo{pages}{113006} (\bibinfo{year}{2008}{\natexlab{b}}).

\bibitem[{\citenamefont{Aoyama et~al.}(2010{\natexlab{a}})\citenamefont{Aoyama,
  Asano, Hayakawa, Kinoshita, Nio, and Watanabe}}]{Aoyama:2010yt}
\bibinfo{author}{\bibfnamefont{T.}~\bibnamefont{Aoyama}},
  \bibinfo{author}{\bibfnamefont{K.}~\bibnamefont{Asano}},
  \bibinfo{author}{\bibfnamefont{M.}~\bibnamefont{Hayakawa}},
  \bibinfo{author}{\bibfnamefont{T.}~\bibnamefont{Kinoshita}},
  \bibinfo{author}{\bibfnamefont{M.}~\bibnamefont{Nio}}, \bibnamefont{and}
  \bibinfo{author}{\bibfnamefont{N.}~\bibnamefont{Watanabe}},
  \bibinfo{journal}{Phys. Rev. D} \textbf{\bibinfo{volume}{81}},
  \bibinfo{pages}{053009} (\bibinfo{year}{2010}{\natexlab{a}}).

\bibitem[{\citenamefont{Aoyama et~al.}(2010{\natexlab{b}})\citenamefont{Aoyama,
  Hayakawa, Kinoshita, and Nio}}]{Aoyama:2010pk}
\bibinfo{author}{\bibfnamefont{T.}~\bibnamefont{Aoyama}},
  \bibinfo{author}{\bibfnamefont{M.}~\bibnamefont{Hayakawa}},
  \bibinfo{author}{\bibfnamefont{T.}~\bibnamefont{Kinoshita}},
  \bibnamefont{and} \bibinfo{author}{\bibfnamefont{M.}~\bibnamefont{Nio}},
  \bibinfo{journal}{Phys. Rev. D} \textbf{\bibinfo{volume}{82}},
  \bibinfo{pages}{113004} (\bibinfo{year}{2010}{\natexlab{b}}).

\bibitem[{\citenamefont{Aoyama et~al.}(2011{\natexlab{a}})\citenamefont{Aoyama,
  Hayakawa, Kinoshita, and Nio}}]{Aoyama:2010zp}
\bibinfo{author}{\bibfnamefont{T.}~\bibnamefont{Aoyama}},
  \bibinfo{author}{\bibfnamefont{M.}~\bibnamefont{Hayakawa}},
  \bibinfo{author}{\bibfnamefont{T.}~\bibnamefont{Kinoshita}},
  \bibnamefont{and} \bibinfo{author}{\bibfnamefont{M.}~\bibnamefont{Nio}},
  \bibinfo{journal}{Phys. Rev. D} \textbf{\bibinfo{volume}{83}},
  \bibinfo{pages}{053003} (\bibinfo{year}{2011}{\natexlab{a}}).

\bibitem[{\citenamefont{Aoyama et~al.}(2011{\natexlab{b}})\citenamefont{Aoyama,
  Hayakawa, Kinoshita, and Nio}}]{Aoyama:2011rm}
\bibinfo{author}{\bibfnamefont{T.}~\bibnamefont{Aoyama}},
  \bibinfo{author}{\bibfnamefont{M.}~\bibnamefont{Hayakawa}},
  \bibinfo{author}{\bibfnamefont{T.}~\bibnamefont{Kinoshita}},
  \bibnamefont{and} \bibinfo{author}{\bibfnamefont{M.}~\bibnamefont{Nio}},
  \bibinfo{journal}{Phys. Rev. D} \textbf{\bibinfo{volume}{83}},
  \bibinfo{pages}{053002} (\bibinfo{year}{2011}{\natexlab{b}}).

\bibitem[{\citenamefont{Aoyama et~al.}(2011{\natexlab{c}})\citenamefont{Aoyama,
  Hayakawa, Kinoshita, and Nio}}]{Aoyama:2011zy}
\bibinfo{author}{\bibfnamefont{T.}~\bibnamefont{Aoyama}},
  \bibinfo{author}{\bibfnamefont{M.}~\bibnamefont{Hayakawa}},
  \bibinfo{author}{\bibfnamefont{T.}~\bibnamefont{Kinoshita}},
  \bibnamefont{and} \bibinfo{author}{\bibfnamefont{M.}~\bibnamefont{Nio}},
  \bibinfo{journal}{Phys. Rev. D} \textbf{\bibinfo{volume}{84}},
  \bibinfo{pages}{053003} (\bibinfo{year}{2011}{\natexlab{c}}).

\bibitem[{\citenamefont{Aoyama et~al.}(2012{\natexlab{a}})\citenamefont{Aoyama,
  Hayakawa, Kinoshita, and Nio}}]{Aoyama:2011dy}
\bibinfo{author}{\bibfnamefont{T.}~\bibnamefont{Aoyama}},
  \bibinfo{author}{\bibfnamefont{M.}~\bibnamefont{Hayakawa}},
  \bibinfo{author}{\bibfnamefont{T.}~\bibnamefont{Kinoshita}},
  \bibnamefont{and} \bibinfo{author}{\bibfnamefont{M.}~\bibnamefont{Nio}},
  \bibinfo{journal}{Phys. Rev. D} \textbf{\bibinfo{volume}{85}},
  \bibinfo{pages}{033007} (\bibinfo{year}{2012}{\natexlab{a}}).

\bibitem[{\citenamefont{Aoyama et~al.}(2012{\natexlab{b}})\citenamefont{Aoyama,
  Hayakawa, Kinoshita, and Nio}}]{Aoyama:2012fc}
\bibinfo{author}{\bibfnamefont{T.}~\bibnamefont{Aoyama}},
  \bibinfo{author}{\bibfnamefont{M.}~\bibnamefont{Hayakawa}},
  \bibinfo{author}{\bibfnamefont{T.}~\bibnamefont{Kinoshita}},
  \bibnamefont{and} \bibinfo{author}{\bibfnamefont{M.}~\bibnamefont{Nio}},
  \bibinfo{journal}{Phys. Rev. D} \textbf{\bibinfo{volume}{85}},
  \bibinfo{pages}{093013} (\bibinfo{year}{2012}{\natexlab{b}}).

\bibitem[{\citenamefont{Aoyama et~al.}(2012{\natexlab{c}})\citenamefont{Aoyama,
  Hayakawa, Kinoshita, and Nio}}]{ae10:PRL}
\bibinfo{author}{\bibfnamefont{T.}~\bibnamefont{Aoyama}},
  \bibinfo{author}{\bibfnamefont{M.}~\bibnamefont{Hayakawa}},
  \bibinfo{author}{\bibfnamefont{T.}~\bibnamefont{Kinoshita}},
  \bibnamefont{and} \bibinfo{author}{\bibfnamefont{M.}~\bibnamefont{Nio}},
  \bibinfo{journal}{Phys.Rev.Lett.} \textbf{\bibinfo{volume}{109}},
  \bibinfo{pages}{111807} (\bibinfo{year}{2012}{\natexlab{c}}).

\bibitem[{\citenamefont{Baikov et~al.}(2013)\citenamefont{Baikov, Maier, and
  Marquard}}]{Baikov:2013ula}
\bibinfo{author}{\bibfnamefont{P.~A.} \bibnamefont{Baikov}},
  \bibinfo{author}{\bibfnamefont{A.}~\bibnamefont{Maier}}, \bibnamefont{and}
  \bibinfo{author}{\bibfnamefont{P.}~\bibnamefont{Marquard}},
  \bibinfo{journal}{Nucl. Phys.} \textbf{\bibinfo{volume}{B877}},
  \bibinfo{pages}{647} (\bibinfo{year}{2013}).

\bibitem[{\citenamefont{Petermann}(1957)}]{Petermann:1957}
\bibinfo{author}{\bibfnamefont{A.}~\bibnamefont{Petermann}},
  \bibinfo{journal}{Helv. Phys. Acta} \textbf{\bibinfo{volume}{30}},
  \bibinfo{pages}{407} (\bibinfo{year}{1957}).

\bibitem[{\citenamefont{Sommerfield}(1958)}]{Sommerfield:1958}
\bibinfo{author}{\bibfnamefont{C.~M.} \bibnamefont{Sommerfield}},
  \bibinfo{journal}{Ann. Phys. (N.Y.)} \textbf{\bibinfo{volume}{5}},
  \bibinfo{pages}{26} (\bibinfo{year}{1958}).

\bibitem[{\citenamefont{Laporta and Remiddi}(1996)}]{Laporta:1996mq}
\bibinfo{author}{\bibfnamefont{S.}~\bibnamefont{Laporta}} \bibnamefont{and}
  \bibinfo{author}{\bibfnamefont{E.}~\bibnamefont{Remiddi}},
  \bibinfo{journal}{Phys. Lett.} \textbf{\bibinfo{volume}{B379}},
  \bibinfo{pages}{283} (\bibinfo{year}{1996}).

\bibitem[{\citenamefont{Kinoshita}(1995)}]{Kinoshita:1995ym}
\bibinfo{author}{\bibfnamefont{T.}~\bibnamefont{Kinoshita}},
  \bibinfo{journal}{Phys. Rev. Lett.} \textbf{\bibinfo{volume}{75}},
  \bibinfo{pages}{4728} (\bibinfo{year}{1995}).

\bibitem[{\citenamefont{Melnikov and van Ritbergen}(2000)}]{Melnikov:1999xp}
\bibinfo{author}{\bibfnamefont{K.}~\bibnamefont{Melnikov}} \bibnamefont{and}
  \bibinfo{author}{\bibfnamefont{T.}~\bibnamefont{van Ritbergen}},
  \bibinfo{journal}{Phys. Rev. Lett.} \textbf{\bibinfo{volume}{84}},
  \bibinfo{pages}{1673} (\bibinfo{year}{2000}).

\bibitem[{\citenamefont{Laporta}(2017)}]{Laporta:2017okg}
\bibinfo{author}{\bibfnamefont{S.}~\bibnamefont{Laporta}},
  \bibinfo{journal}{Phys. Lett.} \textbf{\bibinfo{volume}{B772}},
  \bibinfo{pages}{232} (\bibinfo{year}{2017}).

\bibitem[{\citenamefont{Kinoshita and Lindquist}(1981)}]{Kinoshita:1981vs}
\bibinfo{author}{\bibfnamefont{T.}~\bibnamefont{Kinoshita}} \bibnamefont{and}
  \bibinfo{author}{\bibfnamefont{W.~B.} \bibnamefont{Lindquist}},
  \bibinfo{journal}{Phys. Rev. Lett.} \textbf{\bibinfo{volume}{47}},
  \bibinfo{pages}{1573} (\bibinfo{year}{1981}).

\bibitem[{\citenamefont{Aoyama et~al.}(2015)\citenamefont{Aoyama, Hayakawa,
  Kinoshita, and Nio}}]{Aoyama:2014sxa}
\bibinfo{author}{\bibfnamefont{T.}~\bibnamefont{Aoyama}},
  \bibinfo{author}{\bibfnamefont{M.}~\bibnamefont{Hayakawa}},
  \bibinfo{author}{\bibfnamefont{T.}~\bibnamefont{Kinoshita}},
  \bibnamefont{and} \bibinfo{author}{\bibfnamefont{M.}~\bibnamefont{Nio}},
  \bibinfo{journal}{Phys. Rev.} \textbf{\bibinfo{volume}{D91}},
  \bibinfo{pages}{033006} (\bibinfo{year}{2015}), \bibinfo{note}{[Erratum:
  Phys. Rev. D96, 019901(E)(2017)]}.

\bibitem[{\citenamefont{Marquard et~al.}(2017)\citenamefont{Marquard, Smirnov,
  Smirnov, Steinhauser, and Wellmann}}]{Marquard:2017iib}
\bibinfo{author}{\bibfnamefont{P.}~\bibnamefont{Marquard}},
  \bibinfo{author}{\bibfnamefont{A.~V.} \bibnamefont{Smirnov}},
  \bibinfo{author}{\bibfnamefont{V.~A.} \bibnamefont{Smirnov}},
  \bibinfo{author}{\bibfnamefont{M.}~\bibnamefont{Steinhauser}},
  \bibnamefont{and} \bibinfo{author}{\bibfnamefont{D.}~\bibnamefont{Wellmann}},
  in \emph{\bibinfo{booktitle}{{International Workshop on e+e- Collisions from
  Phi to Psi (PHIPSI17) Mainz, Germany, June 26-29, 2017}}}
  (\bibinfo{year}{2017}), \eprint{arXiv:1708.07138 [hep-ph]},
  \urlprefix\url{http://inspirehep.net/record/1618753/files/arXiv:1708.07138.pdf}.

\bibitem[{\citenamefont{Volkov}(2017)}]{Volkov:2017xaq}
\bibinfo{author}{\bibfnamefont{S.}~\bibnamefont{Volkov}},
  \bibinfo{journal}{Phys. Rev.} \textbf{\bibinfo{volume}{D96}},
  \bibinfo{pages}{096018} (\bibinfo{year}{2017}).

\bibitem[{\citenamefont{Caffo et~al.}(1978)\citenamefont{Caffo, Turrini, and
  Remiddi}}]{Caffo:1978mg}
\bibinfo{author}{\bibfnamefont{M.}~\bibnamefont{Caffo}},
  \bibinfo{author}{\bibfnamefont{S.}~\bibnamefont{Turrini}}, \bibnamefont{and}
  \bibinfo{author}{\bibfnamefont{E.}~\bibnamefont{Remiddi}},
  \bibinfo{journal}{Nucl. Phys.} \textbf{\bibinfo{volume}{B141}},
  \bibinfo{pages}{302} (\bibinfo{year}{1978}).

\bibitem[{\citenamefont{Lepage}(1978)}]{Lepage:1977sw}
\bibinfo{author}{\bibfnamefont{G.~P.} \bibnamefont{Lepage}},
  \bibinfo{journal}{J. Comput. Phys.} \textbf{\bibinfo{volume}{27}},
  \bibinfo{pages}{192} (\bibinfo{year}{1978}).

\bibitem[{\citenamefont{Jegerlehner}(2017{\natexlab{a}})}]{Jegerlehner:2017zsb}
\bibinfo{author}{\bibfnamefont{F.}~\bibnamefont{Jegerlehner}},
  \eprint{arXiv:1711.06089 [hep-ph]}.

\bibitem[{\citenamefont{Ablikim et~al.}(2016)}]{BESIII:2015orh}
\bibinfo{author}{\bibfnamefont{M.}~\bibnamefont{Ablikim}} \bibnamefont{et~al.}
  (\bibinfo{collaboration}{BESIII}), \bibinfo{journal}{Phys. Lett.}
  \textbf{\bibinfo{volume}{B753}}, \bibinfo{pages}{629} (\bibinfo{year}{2016}).

\bibitem[{\citenamefont{Anashin et~al.}(2017)}]{Anashin:2016hmv}
\bibinfo{author}{\bibfnamefont{V.~V.} \bibnamefont{Anashin}}
  \bibnamefont{et~al.}, \bibinfo{journal}{Phys. Lett.}
  \textbf{\bibinfo{volume}{B770}}, \bibinfo{pages}{174} (\bibinfo{year}{2017}).

\bibitem[{\citenamefont{Nomura and Teubner}(2012)}]{Nomura:2012sb}
\bibinfo{author}{\bibfnamefont{D.}~\bibnamefont{Nomura}} \bibnamefont{and}
  \bibinfo{author}{\bibfnamefont{T.}~\bibnamefont{Teubner}},
  \eprint{arXiv:1208.4194 [hep-ph]}.

\bibitem[{\citenamefont{Taylor and Jegerlehner}(2017)}]{BarryTaylor2017}
\bibinfo{author}{\bibfnamefont{B.}~\bibnamefont{Taylor}} \bibnamefont{and}
  \bibinfo{author}{\bibfnamefont{F.}~\bibnamefont{Jegerlehner}}
  (\bibinfo{year}{2017}), \bibinfo{note}{private communication}.

\bibitem[{\citenamefont{Mohr et~al.}(2016)\citenamefont{Mohr, Newell, and
  Taylor}}]{Mohr:2015ccw}
\bibinfo{author}{\bibfnamefont{P.~J.} \bibnamefont{Mohr}},
  \bibinfo{author}{\bibfnamefont{D.~B.} \bibnamefont{Newell}},
  \bibnamefont{and} \bibinfo{author}{\bibfnamefont{B.~N.}
  \bibnamefont{Taylor}}, \bibinfo{journal}{Rev. Mod. Phys.}
  \textbf{\bibinfo{volume}{88}}, \bibinfo{pages}{035009}
  (\bibinfo{year}{2016}).

\bibitem[{\citenamefont{Mohr et~al.}(2017)\citenamefont{Mohr, Newell, Taylor,
  and Tiesinga}}]{Mohr:2017}
\bibinfo{author}{\bibfnamefont{P.~J.} \bibnamefont{Mohr}},
  \bibinfo{author}{\bibfnamefont{D.~B.} \bibnamefont{Newell}},
  \bibinfo{author}{\bibfnamefont{B.~N.} \bibnamefont{Taylor}},
  \bibnamefont{and} \bibinfo{author}{\bibfnamefont{E.}~\bibnamefont{Tiesinga}},
  \bibinfo{journal}{Metrologia}  (\bibinfo{year}{2017}), \bibinfo{note}{in
  press}, \urlprefix\url{https://doi.org/10.1088/1681-7575/aa99bc}.

\bibitem[{\citenamefont{Bouchendira et~al.}(2011)\citenamefont{Bouchendira,
  Clade, Guellati-Khelifa, Nez, and Biraben}}]{Bouchendira:2010es}
\bibinfo{author}{\bibfnamefont{R.}~\bibnamefont{Bouchendira}},
  \bibinfo{author}{\bibfnamefont{P.}~\bibnamefont{Clade}},
  \bibinfo{author}{\bibfnamefont{S.}~\bibnamefont{Guellati-Khelifa}},
  \bibinfo{author}{\bibfnamefont{F.}~\bibnamefont{Nez}}, \bibnamefont{and}
  \bibinfo{author}{\bibfnamefont{F.}~\bibnamefont{Biraben}},
  \bibinfo{journal}{Phys. Rev. Lett.} \textbf{\bibinfo{volume}{106}},
  \bibinfo{pages}{080801} (\bibinfo{year}{2011}).

\bibitem[{\citenamefont{Audi et~al.}(2012)\citenamefont{Audi, Wang, Wapstra,
  Kondev, MacCormick, Xu, and Pfeiffer}}]{AtomicMassRb2012a}
\bibinfo{author}{\bibfnamefont{G.}~\bibnamefont{Audi}},
  \bibinfo{author}{\bibfnamefont{M.}~\bibnamefont{Wang}},
  \bibinfo{author}{\bibfnamefont{A.}~\bibnamefont{Wapstra}},
  \bibinfo{author}{\bibfnamefont{F.}~\bibnamefont{Kondev}},
  \bibinfo{author}{\bibfnamefont{M.}~\bibnamefont{MacCormick}},
  \bibinfo{author}{\bibfnamefont{X.}~\bibnamefont{Xu}}, \bibnamefont{and}
  \bibinfo{author}{\bibfnamefont{B.}~\bibnamefont{Pfeiffer}},
  \bibinfo{journal}{Chinese Physics C} \textbf{\bibinfo{volume}{36}},
  \bibinfo{pages}{1287} (\bibinfo{year}{2012}),
  \urlprefix\url{http://stacks.iop.org/1674-1137/36/i=12/a=002}.

\bibitem[{\citenamefont{Wang et~al.}(2012)\citenamefont{Wang, Audi, Wapstra,
  Kondev, MacCormick, Xu, and Pfeiffer}}]{AtomicMassRb2012b}
\bibinfo{author}{\bibfnamefont{M.}~\bibnamefont{Wang}},
  \bibinfo{author}{\bibfnamefont{G.}~\bibnamefont{Audi}},
  \bibinfo{author}{\bibfnamefont{A.}~\bibnamefont{Wapstra}},
  \bibinfo{author}{\bibfnamefont{F.}~\bibnamefont{Kondev}},
  \bibinfo{author}{\bibfnamefont{M.}~\bibnamefont{MacCormick}},
  \bibinfo{author}{\bibfnamefont{X.}~\bibnamefont{Xu}}, \bibnamefont{and}
  \bibinfo{author}{\bibfnamefont{B.}~\bibnamefont{Pfeiffer}},
  \bibinfo{journal}{Chinese Physics C} \textbf{\bibinfo{volume}{36}},
  \bibinfo{pages}{1603} (\bibinfo{year}{2012}),
  \urlprefix\url{http://stacks.iop.org/1674-1137/36/i=12/a=003}.

\bibitem[{\citenamefont{Mohr et~al.}(2012)\citenamefont{Mohr, Taylor, and
  Newell}}]{Mohr:2012tt}
\bibinfo{author}{\bibfnamefont{P.~J.} \bibnamefont{Mohr}},
  \bibinfo{author}{\bibfnamefont{B.~N.} \bibnamefont{Taylor}},
  \bibnamefont{and} \bibinfo{author}{\bibfnamefont{D.~B.}
  \bibnamefont{Newell}}, \bibinfo{journal}{Rev. Mod. Phys.}
  \textbf{\bibinfo{volume}{84}}, \bibinfo{pages}{1527} (\bibinfo{year}{2012}).

\bibitem[{\citenamefont{Cvitanovi\'c and
  Kinoshita}(1974{\natexlab{a}})}]{Cvitanovic:1974uf}
\bibinfo{author}{\bibfnamefont{P.}~\bibnamefont{Cvitanovi\'c}}
  \bibnamefont{and}
  \bibinfo{author}{\bibfnamefont{T.}~\bibnamefont{Kinoshita}},
  \bibinfo{journal}{Phys. Rev. D} \textbf{\bibinfo{volume}{10}},
  \bibinfo{pages}{3978} (\bibinfo{year}{1974}{\natexlab{a}}).

\bibitem[{\citenamefont{Kinoshita}(1990)}]{Kinoshita:1990}
\bibinfo{author}{\bibfnamefont{T.}~\bibnamefont{Kinoshita}}, in
  \emph{\bibinfo{booktitle}{Quantum electrodynamics}}, edited by
  \bibinfo{editor}{\bibfnamefont{T.}~\bibnamefont{Kinoshita}}
  (\bibinfo{publisher}{World Scientific, Singapore}, \bibinfo{year}{1990}), pp.
  \bibinfo{pages}{218--321}.

\bibitem[{\citenamefont{Aoyama et~al.}(2006)\citenamefont{Aoyama, Hayakawa,
  Kinoshita, and Nio}}]{Aoyama:2005kf}
\bibinfo{author}{\bibfnamefont{T.}~\bibnamefont{Aoyama}},
  \bibinfo{author}{\bibfnamefont{M.}~\bibnamefont{Hayakawa}},
  \bibinfo{author}{\bibfnamefont{T.}~\bibnamefont{Kinoshita}},
  \bibnamefont{and} \bibinfo{author}{\bibfnamefont{M.}~\bibnamefont{Nio}},
  \bibinfo{journal}{Nucl. Phys.} \textbf{\bibinfo{volume}{B740}},
  \bibinfo{pages}{138} (\bibinfo{year}{2006}).

\bibitem[{\citenamefont{Zimmermann}(1969)}]{Zimmermann:1969}
\bibinfo{author}{\bibfnamefont{W.}~\bibnamefont{Zimmermann}},
  \bibinfo{journal}{Commun. Math. Phys.} \textbf{\bibinfo{volume}{15}},
  \bibinfo{pages}{208} (\bibinfo{year}{1969}).

\bibitem[{\citenamefont{Cvitanovi\'c and
  Kinoshita}(1974{\natexlab{b}})}]{Cvitanovic:1974sv}
\bibinfo{author}{\bibfnamefont{P.}~\bibnamefont{Cvitanovi\'c}}
  \bibnamefont{and}
  \bibinfo{author}{\bibfnamefont{T.}~\bibnamefont{Kinoshita}},
  \bibinfo{journal}{Phys. Rev. D} \textbf{\bibinfo{volume}{10}},
  \bibinfo{pages}{3991} (\bibinfo{year}{1974}{\natexlab{b}}).

\bibitem[{\citenamefont{Aoyama et~al.}(2008{\natexlab{c}})\citenamefont{Aoyama,
  Hayakawa, Kinoshita, and Nio}}]{Aoyama:2007bs}
\bibinfo{author}{\bibfnamefont{T.}~\bibnamefont{Aoyama}},
  \bibinfo{author}{\bibfnamefont{M.}~\bibnamefont{Hayakawa}},
  \bibinfo{author}{\bibfnamefont{T.}~\bibnamefont{Kinoshita}},
  \bibnamefont{and} \bibinfo{author}{\bibfnamefont{M.}~\bibnamefont{Nio}},
  \bibinfo{journal}{Nucl. Phys.} \textbf{\bibinfo{volume}{B796}},
  \bibinfo{pages}{184} (\bibinfo{year}{2008}{\natexlab{c}}).

\bibitem[{\citenamefont{Hida et~al.}(2001)\citenamefont{Hida, Li, and
  Bailey}}]{Hida:2001algorithms}
\bibinfo{author}{\bibfnamefont{Y.}~\bibnamefont{Hida}},
  \bibinfo{author}{\bibfnamefont{X.~S.} \bibnamefont{Li}}, \bibnamefont{and}
  \bibinfo{author}{\bibfnamefont{D.~H.} \bibnamefont{Bailey}}, in
  \emph{\bibinfo{booktitle}{Computer Arithmetic, 2001. Proceedings. 15th IEEE
  Symposium on}} (\bibinfo{organization}{IEEE}, \bibinfo{year}{2001}), pp.
  \bibinfo{pages}{155--162}.

\bibitem[{\citenamefont{Bennett et~al.}(2006)}]{Bennett:2006fi}
\bibinfo{author}{\bibfnamefont{G.~W.} \bibnamefont{Bennett}}
  \bibnamefont{et~al.} (\bibinfo{collaboration}{Muon g-2}),
  \bibinfo{journal}{Phys. Rev.} \textbf{\bibinfo{volume}{D73}},
  \bibinfo{pages}{072003} (\bibinfo{year}{2006}).

\bibitem[{\citenamefont{Grange et~al.}(2015)}]{Grange:2015fou}
\bibinfo{author}{\bibfnamefont{J.}~\bibnamefont{Grange}} \bibnamefont{et~al.}
  (\bibinfo{collaboration}{Muon g-2}), \eprint{arXiv:1501.06858
  [physics.ins-det]}.

\bibitem[{\citenamefont{Chapelain}(2017)}]{Chapelain:2017syu}
\bibinfo{author}{\bibfnamefont{A.}~\bibnamefont{Chapelain}}
  (\bibinfo{collaboration}{Muon g-2}), \bibinfo{journal}{EPJ Web Conf.}
  \textbf{\bibinfo{volume}{137}}, \bibinfo{pages}{08001}
  (\bibinfo{year}{2017}).

\bibitem[{\citenamefont{Iinuma}(2011)}]{Iinuma:2011zz}
\bibinfo{author}{\bibfnamefont{H.}~\bibnamefont{Iinuma}}
  (\bibinfo{collaboration}{J-PARC New g-2/EDM experiment Collaboration}),
  \bibinfo{journal}{J. Phys. Conf. Ser.} \textbf{\bibinfo{volume}{295}},
  \bibinfo{pages}{012032} (\bibinfo{year}{2011}).

\bibitem[{\citenamefont{Otani}(2015)}]{Otani:2015jra}
\bibinfo{author}{\bibfnamefont{M.}~\bibnamefont{Otani}}
  (\bibinfo{collaboration}{E34}), \bibinfo{journal}{JPS Conf. Proc.}
  \textbf{\bibinfo{volume}{8}}, \bibinfo{pages}{025008} (\bibinfo{year}{2015}).

\bibitem[{\citenamefont{Kinoshita and Nio}(2004)}]{Kinoshita:2004wi}
\bibinfo{author}{\bibfnamefont{T.}~\bibnamefont{Kinoshita}} \bibnamefont{and}
  \bibinfo{author}{\bibfnamefont{M.}~\bibnamefont{Nio}},
  \bibinfo{journal}{Phys. Rev. D} \textbf{\bibinfo{volume}{70}},
  \bibinfo{pages}{113001} (\bibinfo{year}{2004}).

\bibitem[{\citenamefont{Aoyama et~al.}(2012{\natexlab{d}})\citenamefont{Aoyama,
  Hayakawa, Kinoshita, and Nio}}]{amu10:PRL}
\bibinfo{author}{\bibfnamefont{T.}~\bibnamefont{Aoyama}},
  \bibinfo{author}{\bibfnamefont{M.}~\bibnamefont{Hayakawa}},
  \bibinfo{author}{\bibfnamefont{T.}~\bibnamefont{Kinoshita}},
  \bibnamefont{and} \bibinfo{author}{\bibfnamefont{M.}~\bibnamefont{Nio}},
  \bibinfo{journal}{Phys.Rev.Lett.} \textbf{\bibinfo{volume}{109}},
  \bibinfo{pages}{111808} (\bibinfo{year}{2012}{\natexlab{d}}).

\bibitem[{\citenamefont{Kurz et~al.}(2016)\citenamefont{Kurz, Liu, Marquard,
  Smirnov, Smirnov, and Steinhauser}}]{Kurz:2016bau}
\bibinfo{author}{\bibfnamefont{A.}~\bibnamefont{Kurz}},
  \bibinfo{author}{\bibfnamefont{T.}~\bibnamefont{Liu}},
  \bibinfo{author}{\bibfnamefont{P.}~\bibnamefont{Marquard}},
  \bibinfo{author}{\bibfnamefont{A.~V.} \bibnamefont{Smirnov}},
  \bibinfo{author}{\bibfnamefont{V.~A.} \bibnamefont{Smirnov}},
  \bibnamefont{and}
  \bibinfo{author}{\bibfnamefont{M.}~\bibnamefont{Steinhauser}},
  \bibinfo{journal}{Phys. Rev.} \textbf{\bibinfo{volume}{D93}},
  \bibinfo{pages}{053017} (\bibinfo{year}{2016}).

\bibitem[{\citenamefont{Kurz et~al.}(2015)\citenamefont{Kurz, Liu, Marquard,
  Smirnov, Smirnov, and Steinhauser}}]{Kurz:2015bia}
\bibinfo{author}{\bibfnamefont{A.}~\bibnamefont{Kurz}},
  \bibinfo{author}{\bibfnamefont{T.}~\bibnamefont{Liu}},
  \bibinfo{author}{\bibfnamefont{P.}~\bibnamefont{Marquard}},
  \bibinfo{author}{\bibfnamefont{A.~V.} \bibnamefont{Smirnov}},
  \bibinfo{author}{\bibfnamefont{V.~A.} \bibnamefont{Smirnov}},
  \bibnamefont{and}
  \bibinfo{author}{\bibfnamefont{M.}~\bibnamefont{Steinhauser}},
  \bibinfo{journal}{Phys. Rev.} \textbf{\bibinfo{volume}{D92}},
  \bibinfo{pages}{073019} (\bibinfo{year}{2015}).

\bibitem[{\citenamefont{Teubner}(2017)}]{Teubner:2017phipsi}
\bibinfo{author}{\bibfnamefont{T.}~\bibnamefont{Teubner}}
  (\bibinfo{year}{2017}), \bibinfo{note}{international workshop on $e+e-$
  collisions from Phi to Psi 2017, 26-29 June, 2017, Mainz, Germany},
  \urlprefix\url{https://indico.mitp.uni-mainz.de/event/86/contribution/17/material/slides/0.pdf}.

\bibitem[{\citenamefont{Davier et~al.}(2017)\citenamefont{Davier, Hoecker,
  Malaescu, and Zhang}}]{Davier:2017zfy}
\bibinfo{author}{\bibfnamefont{M.}~\bibnamefont{Davier}},
  \bibinfo{author}{\bibfnamefont{A.}~\bibnamefont{Hoecker}},
  \bibinfo{author}{\bibfnamefont{B.}~\bibnamefont{Malaescu}}, \bibnamefont{and}
  \bibinfo{author}{\bibfnamefont{Z.}~\bibnamefont{Zhang}},
  \eprint{arXiv:1706.09436 [hep-ph]}.

\bibitem[{\citenamefont{Chakraborty et~al.}(2017)\citenamefont{Chakraborty,
  Davies, de~Oliveira, Koponen, Lepage, and Van~de
  Water}}]{Chakraborty:2016mwy}
\bibinfo{author}{\bibfnamefont{B.}~\bibnamefont{Chakraborty}},
  \bibinfo{author}{\bibfnamefont{C.~T.~H.} \bibnamefont{Davies}},
  \bibinfo{author}{\bibfnamefont{P.~G.} \bibnamefont{de~Oliveira}},
  \bibinfo{author}{\bibfnamefont{J.}~\bibnamefont{Koponen}},
  \bibinfo{author}{\bibfnamefont{G.~P.} \bibnamefont{Lepage}},
  \bibnamefont{and} \bibinfo{author}{\bibfnamefont{R.~S.} \bibnamefont{Van~de
  Water}}, \bibinfo{journal}{Phys. Rev.} \textbf{\bibinfo{volume}{D96}},
  \bibinfo{pages}{034516} (\bibinfo{year}{2017}).

\bibitem[{\citenamefont{Blum et~al.}(2016)\citenamefont{Blum, Boyle, Izubuchi,
  Jin, J{\"u}ttner, Lehner, Maltman, Marinkovic, Portelli, and
  Spraggs}}]{Blum:2015you}
\bibinfo{author}{\bibfnamefont{T.}~\bibnamefont{Blum}},
  \bibinfo{author}{\bibfnamefont{P.~A.} \bibnamefont{Boyle}},
  \bibinfo{author}{\bibfnamefont{T.}~\bibnamefont{Izubuchi}},
  \bibinfo{author}{\bibfnamefont{L.}~\bibnamefont{Jin}},
  \bibinfo{author}{\bibfnamefont{A.}~\bibnamefont{J{\"u}ttner}},
  \bibinfo{author}{\bibfnamefont{C.}~\bibnamefont{Lehner}},
  \bibinfo{author}{\bibfnamefont{K.}~\bibnamefont{Maltman}},
  \bibinfo{author}{\bibfnamefont{M.}~\bibnamefont{Marinkovic}},
  \bibinfo{author}{\bibfnamefont{A.}~\bibnamefont{Portelli}}, \bibnamefont{and}
  \bibinfo{author}{\bibfnamefont{M.}~\bibnamefont{Spraggs}},
  \bibinfo{journal}{Phys. Rev. Lett.} \textbf{\bibinfo{volume}{116}},
  \bibinfo{pages}{232002} (\bibinfo{year}{2016}).

\bibitem[{\citenamefont{Kurz et~al.}(2014{\natexlab{b}})\citenamefont{Kurz,
  Liu, Marquard, and Steinhauser}}]{Kurz:2014wya}
\bibinfo{author}{\bibfnamefont{A.}~\bibnamefont{Kurz}},
  \bibinfo{author}{\bibfnamefont{T.}~\bibnamefont{Liu}},
  \bibinfo{author}{\bibfnamefont{P.}~\bibnamefont{Marquard}}, \bibnamefont{and}
  \bibinfo{author}{\bibfnamefont{M.}~\bibnamefont{Steinhauser}},
  \bibinfo{journal}{Phys. Lett.} \textbf{\bibinfo{volume}{B734}},
  \bibinfo{pages}{144} (\bibinfo{year}{2014}{\natexlab{b}}).

\bibitem[{\citenamefont{Prades et~al.}(2009)\citenamefont{Prades, de~Rafael,
  and Vainshtein}}]{Prades:2009tw}
\bibinfo{author}{\bibfnamefont{J.}~\bibnamefont{Prades}},
  \bibinfo{author}{\bibfnamefont{E.}~\bibnamefont{de~Rafael}},
  \bibnamefont{and}
  \bibinfo{author}{\bibfnamefont{A.}~\bibnamefont{Vainshtein}}, in
  \emph{\bibinfo{booktitle}{Lepton Dipole Moments}}, edited by
  \bibinfo{editor}{\bibfnamefont{B.~L.} \bibnamefont{Roberts}}
  \bibnamefont{and} \bibinfo{editor}{\bibfnamefont{W.~J.}
  \bibnamefont{Marciano}} (\bibinfo{publisher}{World Scientific, Singapore},
  \bibinfo{year}{2009}), pp. \bibinfo{pages}{303--319}.

\bibitem[{\citenamefont{Jegerlehner and Nyffeler}(2009)}]{Jegerlehner:2009ry}
\bibinfo{author}{\bibfnamefont{F.}~\bibnamefont{Jegerlehner}} \bibnamefont{and}
  \bibinfo{author}{\bibfnamefont{A.}~\bibnamefont{Nyffeler}},
  \bibinfo{journal}{Phys. Rept.} \textbf{\bibinfo{volume}{477}},
  \bibinfo{pages}{1} (\bibinfo{year}{2009}).

\bibitem[{\citenamefont{Bijnens et~al.}(1995)\citenamefont{Bijnens, Pallante,
  and Prades}}]{Bijnens:1995cc}
\bibinfo{author}{\bibfnamefont{J.}~\bibnamefont{Bijnens}},
  \bibinfo{author}{\bibfnamefont{E.}~\bibnamefont{Pallante}}, \bibnamefont{and}
  \bibinfo{author}{\bibfnamefont{J.}~\bibnamefont{Prades}},
  \bibinfo{journal}{Phys. Rev. Lett.} \textbf{\bibinfo{volume}{75}},
  \bibinfo{pages}{1447} (\bibinfo{year}{1995}), \bibinfo{note}{[Erratum: Phys.
  Rev. Lett.75,3781(1995)]}.

\bibitem[{\citenamefont{Bijnens et~al.}(1996)\citenamefont{Bijnens, Pallante,
  and Prades}}]{Bijnens:1995xf}
\bibinfo{author}{\bibfnamefont{J.}~\bibnamefont{Bijnens}},
  \bibinfo{author}{\bibfnamefont{E.}~\bibnamefont{Pallante}}, \bibnamefont{and}
  \bibinfo{author}{\bibfnamefont{J.}~\bibnamefont{Prades}},
  \bibinfo{journal}{Nucl. Phys.} \textbf{\bibinfo{volume}{B474}},
  \bibinfo{pages}{379} (\bibinfo{year}{1996}).

\bibitem[{\citenamefont{Hayakawa et~al.}(1995)\citenamefont{Hayakawa,
  Kinoshita, and Sanda}}]{Hayakawa:1995ps}
\bibinfo{author}{\bibfnamefont{M.}~\bibnamefont{Hayakawa}},
  \bibinfo{author}{\bibfnamefont{T.}~\bibnamefont{Kinoshita}},
  \bibnamefont{and} \bibinfo{author}{\bibfnamefont{A.~I.} \bibnamefont{Sanda}},
  \bibinfo{journal}{Phys. Rev. Lett.} \textbf{\bibinfo{volume}{75}},
  \bibinfo{pages}{790} (\bibinfo{year}{1995}).

\bibitem[{\citenamefont{Hayakawa et~al.}(1996)\citenamefont{Hayakawa,
  Kinoshita, and Sanda}}]{Hayakawa:1996ki}
\bibinfo{author}{\bibfnamefont{M.}~\bibnamefont{Hayakawa}},
  \bibinfo{author}{\bibfnamefont{T.}~\bibnamefont{Kinoshita}},
  \bibnamefont{and} \bibinfo{author}{\bibfnamefont{A.~I.} \bibnamefont{Sanda}},
  \bibinfo{journal}{Phys. Rev.} \textbf{\bibinfo{volume}{D54}},
  \bibinfo{pages}{3137} (\bibinfo{year}{1996}).

\bibitem[{\citenamefont{Hayakawa and Kinoshita}(1998)}]{Hayakawa:1997rq}
\bibinfo{author}{\bibfnamefont{M.}~\bibnamefont{Hayakawa}} \bibnamefont{and}
  \bibinfo{author}{\bibfnamefont{T.}~\bibnamefont{Kinoshita}},
  \bibinfo{journal}{Phys. Rev.} \textbf{\bibinfo{volume}{D57}},
  \bibinfo{pages}{465} (\bibinfo{year}{1998}), \bibinfo{note}{[Erratum: Phys.
  Rev.D66,019902(E)(2002)]}.

\bibitem[{\citenamefont{Melnikov and Vainshtein}(2004)}]{Melnikov:2003xd}
\bibinfo{author}{\bibfnamefont{K.}~\bibnamefont{Melnikov}} \bibnamefont{and}
  \bibinfo{author}{\bibfnamefont{A.}~\bibnamefont{Vainshtein}},
  \bibinfo{journal}{Phys. Rev. D} \textbf{\bibinfo{volume}{70}},
  \bibinfo{pages}{113006} (\bibinfo{year}{2004}).

\bibitem[{\citenamefont{Bijnens and Prades}(2007)}]{Bijnens:2007pz}
\bibinfo{author}{\bibfnamefont{J.}~\bibnamefont{Bijnens}} \bibnamefont{and}
  \bibinfo{author}{\bibfnamefont{J.}~\bibnamefont{Prades}},
  \bibinfo{journal}{Mod. Phys. Lett.} \textbf{\bibinfo{volume}{A22}},
  \bibinfo{pages}{767} (\bibinfo{year}{2007}).

\bibitem[{\citenamefont{Nyffeler}(2009)}]{Nyffeler:2009tw}
\bibinfo{author}{\bibfnamefont{A.}~\bibnamefont{Nyffeler}},
  \bibinfo{journal}{Phys. Rev. D} \textbf{\bibinfo{volume}{79}},
  \bibinfo{pages}{073012} (\bibinfo{year}{2009}).

\bibitem[{\citenamefont{Pauk and
  Vanderhaeghen}(2014{\natexlab{a}})}]{Pauk:2014rta}
\bibinfo{author}{\bibfnamefont{V.}~\bibnamefont{Pauk}} \bibnamefont{and}
  \bibinfo{author}{\bibfnamefont{M.}~\bibnamefont{Vanderhaeghen}},
  \bibinfo{journal}{Eur. Phys. J.} \textbf{\bibinfo{volume}{C74}},
  \bibinfo{pages}{3008} (\bibinfo{year}{2014}{\natexlab{a}}).

\bibitem[{\citenamefont{G\'{e}rardin et~al.}(2016)\citenamefont{G\'{e}rardin,
  Meyer, and Nyffeler}}]{Gerardin:2016cqj}
\bibinfo{author}{\bibfnamefont{A.}~\bibnamefont{G\'{e}rardin}},
  \bibinfo{author}{\bibfnamefont{H.~B.} \bibnamefont{Meyer}}, \bibnamefont{and}
  \bibinfo{author}{\bibfnamefont{A.}~\bibnamefont{Nyffeler}},
  \bibinfo{journal}{Phys. Rev.} \textbf{\bibinfo{volume}{D94}},
  \bibinfo{pages}{074507} (\bibinfo{year}{2016}).

\bibitem[{\citenamefont{Jegerlehner}(2017{\natexlab{b}})}]{Jegerlehner:2017lbd}
\bibinfo{author}{\bibfnamefont{F.}~\bibnamefont{Jegerlehner}},
  \eprint{arXiv:1705.00263 [hep-ph]}.

\bibitem[{\citenamefont{Colangelo
  et~al.}(2014{\natexlab{a}})\citenamefont{Colangelo, Hoferichter, Nyffeler,
  Passera, and Stoffer}}]{Colangelo:2014qya}
\bibinfo{author}{\bibfnamefont{G.}~\bibnamefont{Colangelo}},
  \bibinfo{author}{\bibfnamefont{M.}~\bibnamefont{Hoferichter}},
  \bibinfo{author}{\bibfnamefont{A.}~\bibnamefont{Nyffeler}},
  \bibinfo{author}{\bibfnamefont{M.}~\bibnamefont{Passera}}, \bibnamefont{and}
  \bibinfo{author}{\bibfnamefont{P.}~\bibnamefont{Stoffer}},
  \bibinfo{journal}{Phys. Lett.} \textbf{\bibinfo{volume}{B735}},
  \bibinfo{pages}{90} (\bibinfo{year}{2014}{\natexlab{a}}).

\bibitem[{\citenamefont{Green et~al.}(2015)\citenamefont{Green, Gryniuk, von
  Hippel, Meyer, and Pascalutsa}}]{Green:2015sra}
\bibinfo{author}{\bibfnamefont{J.}~\bibnamefont{Green}},
  \bibinfo{author}{\bibfnamefont{O.}~\bibnamefont{Gryniuk}},
  \bibinfo{author}{\bibfnamefont{G.}~\bibnamefont{von Hippel}},
  \bibinfo{author}{\bibfnamefont{H.~B.} \bibnamefont{Meyer}}, \bibnamefont{and}
  \bibinfo{author}{\bibfnamefont{V.}~\bibnamefont{Pascalutsa}},
  \bibinfo{journal}{Phys. Rev. Lett.} \textbf{\bibinfo{volume}{115}},
  \bibinfo{pages}{222003} (\bibinfo{year}{2015}).

\bibitem[{\citenamefont{Asmussen et~al.}(2016)\citenamefont{Asmussen, Green,
  Meyer, and Nyffeler}}]{Asmussen:2016lse}
\bibinfo{author}{\bibfnamefont{N.}~\bibnamefont{Asmussen}},
  \bibinfo{author}{\bibfnamefont{J.}~\bibnamefont{Green}},
  \bibinfo{author}{\bibfnamefont{H.~B.} \bibnamefont{Meyer}}, \bibnamefont{and}
  \bibinfo{author}{\bibfnamefont{A.}~\bibnamefont{Nyffeler}},
  \bibinfo{journal}{PoS} \textbf{\bibinfo{volume}{LATTICE2016}},
  \bibinfo{pages}{164} (\bibinfo{year}{2016}).

\bibitem[{\citenamefont{Blum et~al.}(2017{\natexlab{a}})\citenamefont{Blum,
  Christ, Hayakawa, Izubuchi, Jin, Jung, and Lehner}}]{Blum:2016lnc}
\bibinfo{author}{\bibfnamefont{T.}~\bibnamefont{Blum}},
  \bibinfo{author}{\bibfnamefont{N.}~\bibnamefont{Christ}},
  \bibinfo{author}{\bibfnamefont{M.}~\bibnamefont{Hayakawa}},
  \bibinfo{author}{\bibfnamefont{T.}~\bibnamefont{Izubuchi}},
  \bibinfo{author}{\bibfnamefont{L.}~\bibnamefont{Jin}},
  \bibinfo{author}{\bibfnamefont{C.}~\bibnamefont{Jung}}, \bibnamefont{and}
  \bibinfo{author}{\bibfnamefont{C.}~\bibnamefont{Lehner}},
  \bibinfo{journal}{Phys. Rev. Lett.} \textbf{\bibinfo{volume}{118}},
  \bibinfo{pages}{022005} (\bibinfo{year}{2017}{\natexlab{a}}).

\bibitem[{\citenamefont{Blum et~al.}(2017{\natexlab{b}})\citenamefont{Blum,
  Christ, Hayakawa, Izubuchi, Jin, Jung, and Lehner}}]{Blum:2017cer}
\bibinfo{author}{\bibfnamefont{T.}~\bibnamefont{Blum}},
  \bibinfo{author}{\bibfnamefont{N.}~\bibnamefont{Christ}},
  \bibinfo{author}{\bibfnamefont{M.}~\bibnamefont{Hayakawa}},
  \bibinfo{author}{\bibfnamefont{T.}~\bibnamefont{Izubuchi}},
  \bibinfo{author}{\bibfnamefont{L.}~\bibnamefont{Jin}},
  \bibinfo{author}{\bibfnamefont{C.}~\bibnamefont{Jung}}, \bibnamefont{and}
  \bibinfo{author}{\bibfnamefont{C.}~\bibnamefont{Lehner}},
  \bibinfo{journal}{Phys. Rev.} \textbf{\bibinfo{volume}{D96}},
  \bibinfo{pages}{034515} (\bibinfo{year}{2017}{\natexlab{b}}).

\bibitem[{\citenamefont{G\'{e}rardin et~al.}(2017)\citenamefont{G\'{e}rardin,
  Green, Gryniuk, von Hippel, Meyer, Pascalutsa, and
  Wittig}}]{Gerardin:2017ryf}
\bibinfo{author}{\bibfnamefont{A.}~\bibnamefont{G\'{e}rardin}},
  \bibinfo{author}{\bibfnamefont{J.}~\bibnamefont{Green}},
  \bibinfo{author}{\bibfnamefont{O.}~\bibnamefont{Gryniuk}},
  \bibinfo{author}{\bibfnamefont{G.}~\bibnamefont{von Hippel}},
  \bibinfo{author}{\bibfnamefont{H.~B.} \bibnamefont{Meyer}},
  \bibinfo{author}{\bibfnamefont{V.}~\bibnamefont{Pascalutsa}},
  \bibnamefont{and} \bibinfo{author}{\bibfnamefont{H.}~\bibnamefont{Wittig}},
  \eprint{arXiv:1712.00421 [hep-lat]}.

\bibitem[{\citenamefont{Colangelo
  et~al.}(2014{\natexlab{b}})\citenamefont{Colangelo, Hoferichter, Procura, and
  Stoffer}}]{Colangelo:2014dfa}
\bibinfo{author}{\bibfnamefont{G.}~\bibnamefont{Colangelo}},
  \bibinfo{author}{\bibfnamefont{M.}~\bibnamefont{Hoferichter}},
  \bibinfo{author}{\bibfnamefont{M.}~\bibnamefont{Procura}}, \bibnamefont{and}
  \bibinfo{author}{\bibfnamefont{P.}~\bibnamefont{Stoffer}},
  \bibinfo{journal}{JHEP} \textbf{\bibinfo{volume}{09}}, \bibinfo{pages}{091}
  (\bibinfo{year}{2014}{\natexlab{b}}).

\bibitem[{\citenamefont{Pauk and
  Vanderhaeghen}(2014{\natexlab{b}})}]{Pauk:2014rfa}
\bibinfo{author}{\bibfnamefont{V.}~\bibnamefont{Pauk}} \bibnamefont{and}
  \bibinfo{author}{\bibfnamefont{M.}~\bibnamefont{Vanderhaeghen}},
  \bibinfo{journal}{Phys. Rev.} \textbf{\bibinfo{volume}{D90}},
  \bibinfo{pages}{113012} (\bibinfo{year}{2014}{\natexlab{b}}).

\bibitem[{\citenamefont{Colangelo et~al.}(2015)\citenamefont{Colangelo,
  Hoferichter, Procura, and Stoffer}}]{Colangelo:2015ama}
\bibinfo{author}{\bibfnamefont{G.}~\bibnamefont{Colangelo}},
  \bibinfo{author}{\bibfnamefont{M.}~\bibnamefont{Hoferichter}},
  \bibinfo{author}{\bibfnamefont{M.}~\bibnamefont{Procura}}, \bibnamefont{and}
  \bibinfo{author}{\bibfnamefont{P.}~\bibnamefont{Stoffer}},
  \bibinfo{journal}{JHEP} \textbf{\bibinfo{volume}{09}}, \bibinfo{pages}{074}
  (\bibinfo{year}{2015}).

\bibitem[{\citenamefont{Colangelo
  et~al.}(2014{\natexlab{c}})\citenamefont{Colangelo, Hoferichter, Kubis,
  Procura, and Stoffer}}]{Colangelo:2014pva}
\bibinfo{author}{\bibfnamefont{G.}~\bibnamefont{Colangelo}},
  \bibinfo{author}{\bibfnamefont{M.}~\bibnamefont{Hoferichter}},
  \bibinfo{author}{\bibfnamefont{B.}~\bibnamefont{Kubis}},
  \bibinfo{author}{\bibfnamefont{M.}~\bibnamefont{Procura}}, \bibnamefont{and}
  \bibinfo{author}{\bibfnamefont{P.}~\bibnamefont{Stoffer}},
  \bibinfo{journal}{Phys. Lett.} \textbf{\bibinfo{volume}{B738}},
  \bibinfo{pages}{6} (\bibinfo{year}{2014}{\natexlab{c}}).

\bibitem[{\citenamefont{Colangelo et~al.}(2017)\citenamefont{Colangelo,
  Hoferichter, Procura, and Stoffer}}]{Colangelo:2017fiz}
\bibinfo{author}{\bibfnamefont{G.}~\bibnamefont{Colangelo}},
  \bibinfo{author}{\bibfnamefont{M.}~\bibnamefont{Hoferichter}},
  \bibinfo{author}{\bibfnamefont{M.}~\bibnamefont{Procura}}, \bibnamefont{and}
  \bibinfo{author}{\bibfnamefont{P.}~\bibnamefont{Stoffer}},
  \bibinfo{journal}{JHEP} \textbf{\bibinfo{volume}{04}}, \bibinfo{pages}{161}
  (\bibinfo{year}{2017}).

\bibitem[{\citenamefont{Fujikawa et~al.}(1972)\citenamefont{Fujikawa, Lee, and
  Sanda}}]{Fujikawa:1972fe}
\bibinfo{author}{\bibfnamefont{K.}~\bibnamefont{Fujikawa}},
  \bibinfo{author}{\bibfnamefont{B.}~\bibnamefont{Lee}}, \bibnamefont{and}
  \bibinfo{author}{\bibfnamefont{A.}~\bibnamefont{Sanda}},
  \bibinfo{journal}{Phys. Rev. D} \textbf{\bibinfo{volume}{6}},
  \bibinfo{pages}{2923} (\bibinfo{year}{1972}).

\bibitem[{\citenamefont{Czarnecki et~al.}(2003)\citenamefont{Czarnecki,
  Marciano, and Vainshtein}}]{Czarnecki:2002nt}
\bibinfo{author}{\bibfnamefont{A.}~\bibnamefont{Czarnecki}},
  \bibinfo{author}{\bibfnamefont{W.~J.} \bibnamefont{Marciano}},
  \bibnamefont{and}
  \bibinfo{author}{\bibfnamefont{A.}~\bibnamefont{Vainshtein}},
  \bibinfo{journal}{Phys. Rev. D} \textbf{\bibinfo{volume}{67}},
  \bibinfo{pages}{073006} (\bibinfo{year}{2003}), \bibinfo{note}{[Erratum:
  Phys. Rev. D73,119901(E)(2006)]}.

\bibitem[{\citenamefont{Gnendiger et~al.}(2013)\citenamefont{Gnendiger,
  St{\"o}ckinger, and St{\"o}ckinger-Kim}}]{Gnendiger:2013pva}
\bibinfo{author}{\bibfnamefont{C.}~\bibnamefont{Gnendiger}},
  \bibinfo{author}{\bibfnamefont{D.}~\bibnamefont{St{\"o}ckinger}},
  \bibnamefont{and}
  \bibinfo{author}{\bibfnamefont{H.}~\bibnamefont{St{\"o}ckinger-Kim}},
  \bibinfo{journal}{Phys. Rev.} \textbf{\bibinfo{volume}{D88}},
  \bibinfo{pages}{053005} (\bibinfo{year}{2013}).

\end{thebibliography}
